\documentclass[11pt]{article}
\pdfoutput=1 
\usepackage{jheppub}
\usepackage{comment}

\newcommand{\Tr}{\textrm{Tr}}

\newcommand{\bea}{\begin{eqnarray}}
\newcommand{\eea}{\end{eqnarray}}
\newcommand{\bean}{\begin{eqnarray*}}
\newcommand{\eean}{\end{eqnarray*}}

\def\Label#1{\label{#1}%
  \smash{\hbox to0pt{\raise1ex\hbox{\tiny[#1]}\hss}}}
  
\renewcommand{\eqref}[1]{eq.~(\ref{#1})}

\title{A minimal approach to the scattering of physical massless bosons}
\author{Rutger H. Boels}
\emailAdd{Rutger.Boels@desy.de}
\author{and Hui Luo}
\emailAdd{Hui.Luo@desy.de}
\affiliation{II. Institut f\"ur Theoretische Physik, Universit\"at Hamburg\\ Luruper Chaussee 149, D- 22761 Hamburg, Germany }


\abstract{
Tree and loop level scattering amplitudes which involve physical massless bosons are derived directly from physical constraints such as locality, symmetry and unitarity, bypassing path integral constructions. Amplitudes can be projected onto a minimal basis of kinematic factors through linear algebra, by employing four dimensional spinor helicity methods or at its most general using projection techniques. The linear algebra analysis is closely related to amplitude relations, especially the Bern-Carrasco-Johansson relations for gluon amplitudes and the Kawai-Lewellen-Tye relations between gluons and graviton amplitudes. Projection techniques are known to reduce the computation of loop amplitudes with spinning particles to scalar integrals. Unitarity, locality and integration-by-parts identities can then be used to fix complete tree and loop amplitudes efficiently. The loop amplitudes follow algorithmically from the trees. A number of proof-of-concept examples are presented. These include the planar four point two-loop amplitude in pure Yang-Mills theory as well as a range of one loop amplitudes with internal and external scalars, gluons and gravitons. Several interesting features of the results are highlighted, such as the vanishing of certain basis coefficients for gluon and graviton amplitudes. Effective field theories are naturally and efficiently included into the framework. Dimensional regularisation is employed throughout; different regularisation schemes are worked out explicitly. The presented methods appear most powerful in non-supersymmetric theories in cases with relatively few legs, but with potentially many loops. For instance, in the introduced approach iterated unitarity cuts of four point amplitudes for non-supersymmetric gauge and gravity theories can be computed by matrix multiplication, generalising the so-called rung-rule of maximally supersymmetric theories. The philosophy of the approach to kinematics also leads to a technique to control colour quantum numbers of scattering amplitudes with matter, especially efficient in the adjoint and fundamental representations.
}

\begin{document}
\maketitle

\section{Introduction}
Scattering experiments are central to the development of our understanding of the universe at the smallest scales. This physical fact has attracted a large effort to compute observables relevant for these large-scale experiments. Beyond the phenomenological imperative, there is a major formal motivation for computation as well: observables in string and field theories reveal structures of these theories which may not be manifest from their original formulation, spurring development of further insight. These formal and phenomenological motivations have led to many fascinating results in recent years, in a dizzying array of theories, see e.g. \cite{Elvang:2013cua, Henn:2014yza} for an overview. Many of the most powerful results so far are however confined to theories with (highly) unphysical amounts of supersymmetry. While at tree level this is not a major drawback, for quantum corrections there are essential differences between supersymmetric and non-supersymmetric theories. The main purpose of the present paper is to advocate a natural and physical point of view on scattering amplitudes which has ties to many recent developments, is fully compatible with dimensional regularisation but bypasses essential use of supersymmetry. The details of this viewpoint are illustrated here for up to four point amplitudes involving physical massless bosons, but we stress that the methods apply much more generally. 

Perhaps the most basic development this article has ties with are spinor helicity methods. The spinor helicity method in four dimensions for massless theories is the driving force behind compact expressions for many amplitudes. It yields for instance a simple all-multiplicity expression for so-called MHV amplitudes \cite{Parke:1986gb} in Yang-Mills theory. This is still one of the most striking examples of hidden simplicity in complicated Feynman graphs as evidenced by its use in many talks on the subject of scattering amplitudes. In four dimensions for massless particles the physical little group is $SO(2)$ and hence Abelian, which translates to expressions with definite spinor weight for fixed helicity amplitudes. This physical fact drives most of the found simplicity for gluon and scattering amplitudes in four dimensions. Beyond four dimensional massless particles however, spinor helicity methods are not nearly as powerful\footnote{Extensions exist for massive particles in four dimensions \cite{Dittmaier:1998nn}, as well as to higher dimensions, see e.g. \cite{Cheung:2009dc}, \cite{Boels:2012ie} and references therein. See also the very recent  \cite{Arkani-Hamed:2017jhn} for a new approach to massive particles.}. One of the main motivators for this type of extension is dimensional regularisation. Although the particles on the outside of a scattering amplitude can be taken to be four dimensional in an appropriate renormalisation scheme \cite{Bern:1991aq}, this does not hold for internal particles in general. Modern unitarity methods regularly require multi-cuts of loop amplitudes, for which the internal particles are most naturally on the D-dimensional mass-shell. At one loop one can evade this issue in several ways, but at higher loops the problem comes back with a vengeance. Hence methods beyond spinor helicity are required which can treat $D$-dimensional scattering amplitudes at tree and loop level in a natural and most of all efficient way. This article explores a set of such methods, which at loop level has direct connections to earlier work on e.g. Higgs-production amplitudes \cite{Glover:2003cm, Gehrmann:2011aa}, as well as to the computation of form factors \cite{Peskin:1995ev}. 

A second major development this article intersects with is that of amplitude relations: relations among gluon amplitudes as well as relations between gluon and graviton amplitudes. For tree level colour-ordered gluon amplitudes, the first relations in this class were the Kleiss-Kuijff relations \cite{Kleiss:1988ne}. Relatively recently these have been augmented by the Bern-Carrasco-Johannson (BCJ) relations, conjectured in \cite{Bern:2008qj} and proven generically in \cite{BjerrumBohr:2009rd} \cite{Stieberger:2009hq} using string methods and in \cite{Feng:2010my} using purely field theory arguments. The BCJ relations for gluon amplitudes followed from a study of the relation between gravity and gluon amplitudes which traces back to the work of Kawai, Lewellen and Tye (KLT) \cite{Kawai:1985xq} in string theory. The KLT relations are a precise map which constructs tree level gravity scattering amplitudes as a particular sum over products of tree level gauge theory amplitudes with coefficients which are functions of the kinematic invariants. In this article all maps of this type will be referred to as double copy relations. Relations of this type are generic and extend for instance to fermions as well as amplitudes with quite general gluon-graviton-scalar field content \cite{Lancaster:1989qc, Stieberger:2016lng, Du:2017gnh}. The underlying physical mechanism which guarantees relations of the KLT or BCJ variety has remained ill-understood until very recently. Based on the insight that for gluon amplitudes the BCJ relations are a consequence of basic physical constraints such as Poincar\'e symmetry and on-shell gauge invariance \cite{Barreiro:2013dpa}, in \cite{Boels:2016xhc} KLT-type relations for low-multiplicity amplitudes were proven from the same basic principles, and particular counter-examples were obtained. 

The ideas inherent in BCJ and KLT relations have been extended to loop level to an extent at the level of the integrand through a so-called colour-kinematic duality. In brief, if the integrand of a gauge theory amplitude at fixed leg and loop order can be re-written into a form in which the kinematic part of the expressions obey a Lie-algebra-like Jacobi identity, then a gravity amplitude integrand may be constructed through a definite map \cite{Bern:2010ue}. A general correction algorithm starting from a more general gauge theory ansatz has recently yielded the five loop maximal supergravity integrand \cite{Bern:2017yxu}. The problem with this construction is that the colour-kinematic dual form of the integrand at loop level has so far mostly been achieved on a case by case basis. A direct constructive approach to graviton-containing loop amplitudes would therefore be most welcome, especially in non-supersymmetric theories.

A third major development this article connects to is the drive to explore quantum field theory beyond Lagrangians and Lagrangian-based calculation methods such as Feynman graphs. Lagrangian methods and path integrals are very powerful, but inherently suffer from the drawback of having to choose coordinates on field space to path-integrate over. Hence, many different-looking Lagrangians generate the same physics such as that encapsulated in S-matrices. This can hide much of the structure of the theory, such as hidden integrability properties or symmetries. Also, as alluded to above, there are known cases where the result of a computation is much simpler than the computation itself. It is well-known that perturbative unitarity constraints residues at poles and discontinuities at branch cuts of scattering amplitudes. A question is then if complete amplitudes may be constructed from physical requirements such as symmetry and unitarity alone, bypassing Feynman graphs. A prime modern example of such a unitarity-based method is BCFW on-shell recursion \cite{Britto:2004ap, Britto:2005fq} for pure Yang-Mills and gravity theories at tree level. Based on a certain high energy limit of the amplitudes, see. e.g. \cite{ArkaniHamed:2008yf}, one can derive full amplitudes from the residue information at the poles. For extending this philosophy to loop level a large literature exists, see \cite{Elvang:2013cua, Henn:2014yza} for overviews. The philosophy used in this article traces in contrast back to earlier literature, especially \cite{Bern:1994cg}. In essence, one uses unitarity cuts to determine coefficients of certain integral functions. See also  \cite{Arkani-Hamed:2016rak, Rodina:2016mbk} at tree level and \cite{Abreu:2017xsl} as well as \cite{Bern:2017tuc} at loop level for work directly related to but technically distinct from the approach advocated here.

The slogan of the approach advocated in \cite{Boels:2016xhc} and worked out in more detail here is `amplitudes as vectors'. In short, solving the on-shell gauge invariance and symmetry constraints gives all possible tensor structures that can appear in a scattering amplitude with fixed field content. These tensor structures necessarily span a vector space. Every amplitude is a vector in this space: they can be written as a unique linear combination of a set of chosen basis elements for the tensor structures, where the coefficients are functions of Mandelstam invariants only. These linear combinations are closely related to amplitude relations of either BCJ or double copy type at tree level. One goal of this article is to analyse these vector spaces in detail motivated by physical constraints. The power of this approach is the minimality of the assumptions. Hence we can for instance illustrate the uses of such a basis directly in both string and field theory. In particular, our techniques function for quite general effective field theories such as those that arise in the analysis of beyond the standard model physics. A related application of similar technology can be found in for instance  \cite{Barreiro:2013dpa}, \cite{Arkani-Hamed:2016rak} and in \cite{Bern:2017tuc}. 

This article is structured as follows: in section \ref{sec:onshellconstraints} we introduce general concepts and techniques. In section \ref{sec:fourpsingscal} these techniques are applied to a range of four point scattering amplitudes with at least a single scalar, at tree level. This is followed by a section \ref{sec:fourpnoscal}, on amplitudes with four spin one or spin two particles. The focus then turns to one loop level amplitudes in section \ref{sec:oneloop}. The techniques used to compute the two-loop planar four gluon amplitude are presented in section \ref{sec:twoloop4pt}. General observations for multi-loop amplitudes follow in section \ref{sec:morethanoneloop}, and include a discussion of iterated unitarity cuts as well as colour quantum numbers in loops. A discussion and conclusion section rounds of the presentation. In appendix A a dictionary is provided between tensor structure basis and spinor helicity expressions in four dimensions, including an analysis of dimension-induced relations. 
 
\section{Solving the on-shell constraints}\label{sec:onshellconstraints}

\subsection{On-shell constraints}
The on-shell constraints are a sub-set of all physically reasonable demands scattering amplitudes in very general theories in flat space have to obey:
\begin{itemize}
\item Poincar\'e symmetry 
\item transversality
\item on-shell gauge invariance
\end{itemize}
Poincar\'e symmetry leads among other consequences to the well-known Wigner classification of on-shell states \cite{Wigner:1939cj}. That is, every external state (taken to be plane wave) is characterised at minimum by an irreducible representation of the little group and an on-mass-shell momentum. In this article the momenta will be massless unless stated otherwise.  Here we will assume as usual no continuous spin variables play a role. As Coleman and Mandula \cite{Coleman:1967ad} have shown the only bosonic symmetry physically allowed in addition to Poincar\'e symmetry except for conformal symmetry is an internal symmetry, typically characterised by an irreducible representation of a Lie algebra and potentially finite group charges. Poincar\'e symmetry also leads to momentum conservation through Noether's theorem, i.e.
\begin{equation}
\sum_i p_i = 0
\end{equation}
for the inward-pointing momenta convention followed here. Physical (e.g. real and positive) energies summing to zero would set all energies to zero. In order to preserve sufficient generality we will treat momenta without such conditions in this article. Equivalently, we can treat all momenta as complex-valued vectors.

The `embedding tensors' of the little group into the Poincar\'e group play a central role. For bosons these are simply products of photon polarisation vectors, properly projected unto the required irrep. Scattering amplitudes are therefore functions,
\begin{equation}
A_n\left(\{\xi_1, p_1\}, \ldots, \{\xi_n, p_n\} \right)
\end{equation}
where the $\xi$ are general polarisation tensors. Scattering amplitudes transform as little group tensors. For a physicist this simply implies that all space-time indices have to be contracted using the metric.  If a local or global internal symmetry is present, an appropriate irrep index for this symmetry should be added to the set of quantum numbers for each leg. Since the amplitude has to transform as a little group tensor the amplitude is \emph{multilinear} in the polarisation vectors of all the legs: the amplitude to all loop orders simply proportional to the product of all polarisation tensors. In Feynman graph perturbation theory this property is manifest. 

The polarisation tensors obey two types of constraint. First, the constituent photon polarisations are transverse to the momentum in that leg,
\begin{equation}
\xi^i_\mu p_i^{\mu} = 0
\end{equation}
This will be referred to as transversality. Second, for massless spin one or spin two matter Noether's second theorem \cite{Noether:1918zz} dictates that for local gauge or coordinate symmetry replacing a polarisation vector by the momentum in this leg makes the amplitude vanish
\begin{equation}
A\left( \xi^i_{\mu} \rightarrow p^i_{\mu} \right) = 0
\end{equation}
This will be referred to as on-shell gauge invariance.

There are three further, physically reasonable constraints which will be used if concrete physical scattering amplitudes are needed below
\begin{itemize}
\item Bose symmetry
\item locality
\item unitarity
\end{itemize}
as well as a fourth constraint which would be interesting to pursue but that will not be used here,
\begin{itemize}
\item causality
\end{itemize}
For definiteness these constraints will not be included in the term `on-shell constraints' as it will be used in this article, as especially unitarity and locality behave differently depending on loop order. 

Bose symmetry implies that if all the quantum numbers of two indistinguishable, integer spin particles are interchanged the amplitude should remain invariant. The quantum numbers include the momentum, the helicity as well as all internal quantum numbers such as colour or flavour 
\begin{equation}
\{\xi_i, p_i, a_i \} \leftrightarrow \{\xi_j, p_j, a_j \}
\end{equation}
Furthermore, `indistinguishable' particles in this context simply belong to the same helicity and internal symmetry irreps as the swap of indices simply would not make sense otherwise. Bose symmetry naturally intertwines internal and space-time symmetry properties. Fermi symmetry for physical fermions would introduce a minus sign for swaps of indistinguishable particles.

\subsection{Solving the on-shell constraints}
The on-shell constraints can be solved systematically as shown for instance in \cite{Boels:2016xhc}. This will be quickly reviewed here, mostly illustrated by the four point example. The first step is to solve momentum conservation. For four particles the Mandelstam invariants are 
\begin{equation}
s= (p_1 + p_2)^2 \quad t= (p_2 + p_3)^2 \quad u= (p_2 + p_4)^2  
\end{equation}
which obey $s+t+u=\sum_i m_i^2$. Momentum conservation can also be used to eliminate one chosen momentum, say $p_4$, from all contractions with external polarisation vectors, reducing this to contractions with the other momenta. Transversality for $\xi_1,\xi_2$ and $\xi_3$ implies that only contractions with two other momenta are non-zero for each vector. For $\xi_4$ some care has to be taken as $p_4$ has been eliminated by momentum conservation and in expressions one may therefore encounter in the four particle example
\begin{equation}
\ldots \left(p_1 + p_2 + p_3 \right) \cdot \xi_4 = 0
\end{equation}
Hence, for this leg one can always solve for instance $\xi_4 \cdot p_1$ in terms of the other products when writing a full Ansatz up to transversality. With these choices, momentum conservation and transversality have been solved for all external momenta. The order of perturbation theory is used implicitly by excluding tensor integrals where a polarisation contracts with a momentum (e.g. $\xi \cdot l$): the derivation applies after integration\footnote{Although polarisation tensors contracted with loop momenta can be considered similar to the approach in this section, in that case on-shell gauge invariance needs to take into account integration by parts identities. We did not find an efficient method to do this, and we have therefore not pursued this direction. }. 

Having solved Poincar\'e symmetry and transversality one can construct an Ansatz 
\begin{equation}\label{eq:genansatz}
A = \sum_i \alpha_i(s,t) T_i
\end{equation}
where $T_i$ ranges over all possible contractions of the polarisation vectors in the problem with either metrics connecting two polarisations or momenta, subject to the solution to the momentum conservation and transversality constraints. In this article we will only study parity even solutions, i.e. we do not consider the epsilon tensor. Parity-odd tensor structures containing the epsilon symbol solving the on-shell constraints for $n$-particles can be obtained from the parity-even ones for dimension $D$ such that $n>D$, (see \cite{bartelmannthesis}). For the case $n\leq D$, brute force can be applied. In four dimensions, there are for instance $8$ linearly independent parity odd tensor structures for four particles. This number could also have been obtained by counting helicity amplitudes. Parity odd amplitudes are interesting as they connect to anomalies, see e.g. \cite{Chen:2014eva}. 

Naively, on-shell gauge invariance might seem to give only one equation for every photon polarisation vector in the problem. In fact, there there is one equation for every independent tensor structure after replacing the polarisation vector by its momenta. The equations can easily be isolated using the same type solution of momentum conservation and transversality as discussed above, in this case for one polarisation vector less. Given an Ansatz $T_i$, one can therefore derive a system of linear equations for the coefficients $\alpha_i$. Solving this system by standard linear algebra methods then results in a number of independent solutions to the on-shell constraints. Some information on the size of Ans\"atze and sizes of solution spaces is given in \cite{Boels:2016xhc} (see also  \cite{Barreiro:2013dpa}). 

It is natural to treat the coefficients $\alpha_i$ in equation \eqref{eq:genansatz} as the entries of a vector. The on-shell constraints yield in this view a matrix equation which is a function only of the independent Mandelstam invariants, say $s$ and $t$ in the four particle case. The independent solutions to the on-shell constraints then span a vector space where the coefficients are a function of Mandelstams and the dimension. Denote these solutions
\begin{equation}\label{eq:gensol}
B_j = \sum_i \beta_{j,i} (s,t) T_i
\end{equation}
The number of $B$'s is generically much smaller than the number of $T$'s. Independency of the solutions translates into the statement that the vectors $\beta$ are independent,
\begin{equation}
n_j \beta_{j,i}  = 0 \rightarrow n_j = 0
\end{equation}
for all vectors $n_j$ that are functions of Mandelstam invariants. 

Given any other solution to the on-shell constraints such as for instance a certain physical amplitude $A$, there are for a particular chosen basis $B_i$ always coefficients such that
\begin{equation}\label{eq:amplitudesinbasis}
A = \sum_j b_j(s,t) B_j
\end{equation}
One way of computing these coefficients $b_j$ is to express the left-hand and right hand sides in terms of the structures $T_i$ to obtain
\begin{equation}\label{eq:compTmats}
\sum_i \alpha_i T_i =  \sum_{j,i} b_j \beta_{j,i} T_i
\end{equation}
Since the $T_i$ are independent since these solve momentum conservation and transversality, this amounts to a matrix equation
\begin{equation}
\alpha_i = b_j \beta_{j,i}
\end{equation}
which has a solution that can be obtained by simple linear algebra methods: construct the matrix $\{\alpha, \beta_1, \ldots \beta_n\} $ where the individual vectors form the columns. The sought-for relation is than the one-dimensional kernel of this rectangular matrix, typically easily obtained by computer algebra. 

This highly useful technique of computing relations as the kernel of a matrix also works for instance for uncovering relations between amplitudes. If there is a relation between scattering amplitudes where the coefficients are functions of Mandelstam invariants, then there is exactly the same relation between the coefficients of an expansion in a chosen basis, and vice-versa. Hence, after obtaining the coefficients of the basis expansion for all amplitudes in a given set one can systematically construct all independent relations of this type by considering the kernel of the matrix formed out of the expansion coefficients. A drawback of the linear algebra technique is that one needs equation \eqref{eq:compTmats} to be satisfied explicitly. This is for loop amplitudes not manifestly the case, and a different technique must be used here.  

\subsection{P-matrix and dimension dependence}
Given a basis $B_j$ there are other ways to solve equations of the type in equation \eqref{eq:amplitudesinbasis} which give the expressions of a given amplitude $A$ in terms of this basis. One such technique is the construction of a matrix which will be referred to as `P-matrix'. This is essentially equivalent to the approach first advocated in \cite{Glover:2003cm} for certain four point amplitudes. To uncover this matrix, multiply equation \eqref{eq:amplitudesinbasis} by a particular basis element $B_k$ and sum over the helicities of all the particles, 
\begin{equation}\label{eq:P-matrixAction}
\sum_{\textrm{hel}}  \left[B_k A \right] = \sum_j b_j \sum_{\textrm{hel}} \left[B_k B_j \right]
\end{equation}
In the following it will become clear that summing over helicities like this is highly reminiscent an inner product on a vector space. The matrix on the right-hand side will be dubbed the `P-matrix':
\begin{equation}
P_{kj} =\sum_{\textrm{hel}} \left[B_k B_j \right]
\end{equation}
The completeness relation for polarisation sums can be used repeatedly to express the entries of the P-matrix as functions of Mandelstam invariants and the dimension only. Dimension dependence enters through
\begin{equation}
\sum_{\textrm{hel}_i}  \xi_i \cdot \xi_i = D-2
\end{equation}
If the P-matrix can be inverted as a matrix, the linear equation in \eqref{eq:P-matrixAction} has a unique solution for the coefficients $b_j$ which are the sought-for projection coefficients. Note the appearance of the word `if' in the previous sentence. So far the tensor structures $T_i$ and $B_j$ have been assumed to be independent. While this is true in general dimensions, special relations may exist in integer dimensions. In four dimensions for instance five momentum vectors necessarily obey at least a single relation. Working out these Gramm determinant constraints is typically very cumbersome. Here however the P-matrix offers a short-cut: if a linear relation exists in a special dimension for the $B$'s, then the P-matrix develops a kernel, which can be computed. On the other hand, non-linear relations between the tensor structures can be disregarded for our purposes as this would spoil the required linear dependence on polarisation vectors\footnote{We would like to thank Nima Arkani-Hamed for a discussion on this point.}. Vice-versa, a non-trivial kernel of the P-matrix implies that a linear relation between the $B$'s must exist. For special integer dimensions, this provides a concrete tool to compute these linear relations, bypassing Gramm determinants. In the special case of four dimensions these relations can also be uncovered using four dimensional spinor helicity methods as explained in Appendix A.

\subsection{First results: three point amplitudes}
Three point scattering amplitudes are special as by momentum conservation all inner products of momenta are proportional to masses. Here, mostly cases with vanishing external masses are treated for simplicity. It should be stressed that unbroken, $D$ dimensional Poincar\'e symmetry is assumed throughout and no reality conditions are imposed on momenta. 

\subsubsection*{s-s-s}
For three scalar particles there is no gauge invariance to take into account. The amplitude is always proportional to a coupling constant,
\begin{equation}\label{eq:threescalAmp}
B \propto \lambda'
\end{equation}
Assigning the scalar fields a canonical mass-dimension gives the coupling constant $\lambda$ mass dimension $1$. With massive particles and and a consistent massless limit, the most general possibility reads
\begin{equation}
B \propto \lambda_1 m_1 + \lambda_2 m_2 + \lambda_3 m_3 + \lambda_4 
\end{equation}
where $\lambda_i$ has mass dimension zero for $i=1,2,3$ and $1$ for $i=4$.

\subsubsection*{g-s-s}
Solving the on-shell constraints for a general arbitrary spin bosonic field coupled to two scalars is straightforward, see for instance \cite{Boels:2012if}. For two scalars and a single spin-one particle, transversality and momentum conservation lead to an Ansatz,
\begin{equation}
B \sim \xi^3_{\mu} (p_1 - p_2)^{\mu}
\end{equation}
where the gluon is labelled as particle $3$. This expression is on-shell gauge invariant for real mass if and only if the masses of the two scalar particles is the same since
\begin{equation}
(p_1 + p_2) \cdot (p_1 - p_2)  = p_1^2 - p_2^2 = m_1^2 - m_2^2
\end{equation} 
Note that in the case of equal masses the scattering amplitude is anti-symmetric under exchange of the two scalar legs. This will have to be compensated by additional  quantum numbers (such as colour) to obtain Bose-symmetry if the scalars are indistinguishable.

\subsubsection*{G-s-s}
The extension to gravitons follows in this case most easily from considering the extension to the so-called $\mathcal{N}=0$ multiplet. That is, the graviton is part of a reducible multiplet formed by multiplying two photon polarisation vectors,
\begin{equation}
\xi^{I}_{L,\mu} \xi^{J}_{R, \nu} = \xi^{[I}_{\mu} \xi^{J]}_{\nu} + \left(\xi^{\{I}_{\mu} \xi^{J\}}_{\nu} - \frac{\delta^{IJ}}{D-2}\xi^{I}_{\mu} \xi_{I,\nu} \right) +  \frac{\delta^{IJ}}{D-2}\xi^{I}_{\mu} \xi_{I,\nu}
\end{equation}
This multiplet contains  the two-form (anti-symmetric representation), the graviton (traceless symmetric representation) and the trace (the dilaton) respectively. Since this combination of fields is ubiquitous within supergravity theories this reducible, bosonic multiplet is dubbed the $\mathcal{N}=0$ multiplet. In general we will refer to the two photon polarisations as the `left' and `right' polarisations. The Ansatz excludes the direct contraction of the left and right polarisations of a single leg as this can be included by taking the little group trace of the two polarisation vectors. For two scalars, transversality and momentum conservation then lead to a single term Ansatz,
\begin{equation}
B \sim \left( \xi^3_{L,\mu} (p_1 - p_2)^{\mu} \right)\left( \xi^3_{R,\mu} (p_1 - p_2)^{\mu}\right)
\end{equation}
Hence, on-shell gauge invariance only exists for equal mass scalars. Note that for massless scalars, the dilaton and two-form do not couple. This expression is automatically Bose symmetric for the scalar legs. As is well known, this expression is the double copy of two scattering amplitudes with a single gluon and two scalars. 

\subsubsection*{g-g-s}
Two gluons and a scalar lead to a two element Ansatz in this case,
\begin{equation}
T = \{\xi_1 \cdot \xi_2, (p_2 \cdot \xi_1) (p_1 \cdot \xi_2) \}
\end{equation}
that solves momentum conservation and transversality. The second element is on-shell gauge invariant for a massless scalar leg, while the first is not. This leads to one solution to the on-shell constraints in this case. The solution is Bose-symmetric without taking into account colour information.

If the scalar has a mass $m$, the number of solutions to the on-shell constraints is the same, but the explicit expression changes, i.e.
\begin{equation}
B \propto m^2 (\xi_1 \cdot \xi_2 ) - 2 (p_2 \cdot \xi_1) (p_1 \cdot \xi_2) 
\end{equation}
Note that for three massless particles in general the terms in an Ansatz with different numbers of metric contractions cannot mix in the three particle case. With even a single mass this is no longer true, as this example shows clearly. In this particularly simple case, one can identify the corresponding Lagrangian easily: this term is a $\phi \Tr F^2$ interaction term evaluated on-shell. This is part of a larger remark: special solutions to the on-shell constraints are always obtained from Lorentz-invariant contractions of products of linearised field strength tensors.

\subsubsection*{G-G-s}
The Ansatz for the case of a $\mathcal{N}=0$ multiplets on two legs and a scalar contains seven elements. At the outset, it is clear one solution to the on-shell constraints can be obtained by taking a product of two $s-g-g$ solutions, one for the left and one for the right polarisation. This solution also obviously extends to the massive scalar case. 
\begin{equation}
B \propto \left[m^2 (\xi^R_1 \cdot \xi^R_2 ) - 2 (p_2 \cdot \xi^R_1) (p_1 \cdot \xi^R_2) \right] \left[ m^2 (\xi^L_1 \cdot \xi^L_2 ) - 2 (p_2 \cdot \xi^L_1) (p_1 \cdot \xi^L_2) \right] 
\end{equation}
This solution is Bose-symmetric as well as overall left-right symmetric. In the massless scalar case, it is also left-right symmetric in individual graviton legs. 

However, this is not the only solution to the on-shell constraints in the massless scalar case. A second solution exists which features two anti-symmetric particles for a massless scalar which reads 
\begin{multline}
B \propto 2 p_1 \cdot \xi^R_2 ((p_2 \cdot \xi_1^R )(\xi_1^L \cdot \xi_2^L) -  (p_2 \cdot \xi_1^L )(\xi_1^R \cdot \xi_2^L))
    +\\   2 p_1 \cdot \xi_2^L  ((p_2 \cdot \xi_1^L) (\xi_1^R \cdot \xi_2^R) - (p_2 \cdot \xi^R_1) (\xi_1^L \cdot \xi_2^R) )
\end{multline}
This solution features explicit contractions between left and right polarisation vectors and it is Bose-symmetric. The most general solution to the on-shell constraints is a linear combination of this solution and the double-copy-type solution.

\subsubsection*{G-g-s}
For a massless scalar leg, a single gluon and single $\mathcal{N}=0$ multiplet leg a single solution, 
\begin{equation}
B \propto (p_1 \cdot \xi_2)( p_2 \cdot \xi^R_1) (p_2 \cdot \xi^L_1)  
\end{equation}
is obtained. Interestingly, there are no amplitudes of this form with a massive scalar leg, or with a single anti-symmetric form instead of the graviton. 

\subsubsection*{g-g-G}
The Ansatz for  two gluons and one $\mathcal{N}=0$ multiplet contains eight elements. Three solutions to the on-shell constraints exist. The simplest solution contains no metric contractions and reads
\begin{equation}
B_1 \propto (p_2 \cdot \xi^L_1) (p_2 \cdot \xi^R_1) (p_1 \cdot \xi_2)(p_2  \cdot \xi_3)
\end{equation}
The other two solutions are the symmetric and anti-symmetric parts w.r.t. the polarisation vectors of the $\mathcal{N}=0$ multiplet leg of
\begin{equation}
B \propto \left[(p_2 \cdot \xi^R_1  ) (p_2  \cdot \xi_3) (\xi^L_1 \cdot \xi_2) + (p_2 \cdot \xi^R_1  ) (p_1  \cdot \xi_2) (\xi^L_1 \cdot \xi_3 ) - (p_2 \cdot \xi^R_1  ) (p_2  \cdot \xi^L_1) (\xi_2 \cdot \xi_3)  \right]_{S/AS}
\end{equation}
Both these solutions contain a single metric contraction.

\subsubsection*{g-G-G}

The nineteen element Ansatz with two  $\mathcal{N}=0$ multiplets and one gluon gives five independent gauge invariant solutions. As before, there is one solution without any metric contractions, which is
\begin{equation}
B_1 \propto (p_2 \cdot \xi^L_1) (p_2 \cdot \xi^R_1) (p_1 \cdot \xi^L_2)(p_1 \cdot \xi^R_2)(p_2  \cdot \xi_3)
\end{equation}
The other four solutions simply split into the four possibilities of having symmetric/anti-symmetric polarisation tensors for legs one and two. These contain only one metric contraction.

\subsubsection*{g-g-g / G-G-G}
These cases were discussed in some detail in \cite{Boels:2016xhc}. For gluons, two solutions to the on-shell constraints exists of different mass-dimension. One is proportional to the Yang-Mills amplitude,
\begin{equation}
B_1 = -(\xi_1 \cdot \xi_2 )(p_2 \cdot \xi_3)- (\xi_1 \cdot \xi_3 )(p_1 \cdot \xi_2) + (\xi_2 \cdot \xi_3 )(p_2 \cdot \xi_1)
\end{equation}
while the other is generated by the $F^3$ interaction,
\begin{equation}
B_2 = (p_2 \cdot \xi_1)(p_1 \cdot \xi_2)(p_2 \cdot \xi_3) 
\end{equation}
For three  $\mathcal{N}=0$ multiplets, the solutions to the on-shell constraints are always of the double copy type. In general, products of two gluon amplitudes are a special sub-class of solutions to the on-shell constraints for amplitudes involving gravitons. 

\subsection{General construction methods}
Two remarks scattered throughout the previous section deserve amplification as they provide general solution classes to the on-shell constraints. One is that each scalar build out of linearised field strength tensors for each external polarisation vector and momenta obeys the on-shell constraints
\begin{equation}
F^{\textrm{lin}}_{\mu\nu} (\xi_i, p_i) = p_{i, \mu} \xi_{i, \nu} - p_{i, \nu} \xi_{i, \mu}  
\end{equation}
This observation is obviously a part of the connection of the present `on-shell' article to Lagrangian operator methods, which however will not be pursued here. The second remark is that special solutions to the on-shell constraints for a fixed number of points follow from multiplication of previously obtained solutions of this type. The field content follows from the usual tensor product rules. This is the fuel that drives the KLT relations. However, one can also multiply for instance an amplitude with a gluon and three scalars with an amplitude with three gluons and a scalar to create amplitudes with four gluons as 
\begin{equation}
A(\xi_1, \xi_2, \xi_3, \xi_4) = A(\xi_1, s,s,s)A(s, \xi_2, \xi_3, \xi_4) 
\end{equation}
This leads to a bonus third remark: taking momenta collinear on two different gluon-containing amplitudes in principle gives an amplitude with a graviton-like state. This is likely the physical mechanism behind the explicit set of relations obtained in \cite{Stieberger:2016lng}.

\section{Four point amplitudes with one or more scalars}\label{sec:fourpsingscal}

In the previous section a general strategy for solving the on-shell constraints and a series of three point examples were presented. In the present section the strategy will be applied in a number of example four point amplitudes, slowly increasing the complexity. New in the more-than-three-particle case are unitarity constraints due to the non-trivial kinematics. For four particle amplitudes this translates into possible poles in the $s$, $t$ and $u$ channels. Amplitudes cannot have poles in overlapping channels. In this case this implies that the residue at the kinematic poles must be a local function by locality: it cannot contain any further poles. Moreover, the residue must be a product of two three point amplitudes summed over all internal helicities. This will be used below. 

\subsection{Four scalars}
For four scalars without internal symmetry, the scattering amplitude must be a mass-dimension zero, scalar function. The simplest scattering amplitude one can write down is the local, constant amplitude,
\begin{equation}
A_4 = \lambda
\end{equation}
With a dimension-one coupling constant $\lambda'$ and no internal symmetry the most general solution reads
\begin{equation}
A_4 = \lambda + (\lambda')^2 \left(\frac{1}{s} + \frac{1}{t} +\frac{1}{u}  \right)
\end{equation}
The functional form is dictated by unitarity and locality: there can be no higher order poles, or poles in overlapping channels. The latter would be for instance an $s$-channel pole at a $t$ or $u$ channel residue. The residues at the poles are manifestly products of lower point amplitudes\footnote{There is an issue here due to space-time signature and reality conditions, see \cite{Schuster:2008nh} for a discussion.}. Note for instance that taken by itself
\begin{equation}
\frac{1}{s t}
\end{equation}
is ruled out as a scattering amplitude for four scalar particles as it contains overlapping poles. Bose-symmetry fixes the relative coupling constants between the poles, which rules out more complicated combinations of poles. On the poles the residues are fixed in terms of the three scalar amplitude, equation \eqref{eq:threescalAmp}. With internal symmetries, the formulas become more involved.  

\subsection{One gluon and three scalars}
 The basis of tensor structures in the convention introduced above reads
\begin{equation}
T_i = \{ \xi_1 \cdot p_2, \xi_1 \cdot p_3 \}
\end{equation}
and contains $2$ elements. Any amplitude is a linear combination of these two, 
\begin{equation}
B = \vec{\beta} \cdot \vec{T}
\end{equation}
There is one constraint equation on $\vec{\beta}$ derived from on-shell gauge invariance:
\begin{equation}
 \{s, u \} \cdot \vec{\beta} = 0
\end{equation}
A local solution to this constraint is given by the amplitude
\begin{equation}\label{eq:onegluon}
B =  u \, \xi_1 \cdot p_2 - s\,  \xi_1 \cdot p_3
\end{equation}
Local here means an expression manifestly without explicit poles in the Mandelstam invariants. Any amplitude with one gluon and three scalars must be proportional to this one $B$, with the constants of proportionality general functions of the Mandelstam invariants. This basis choice is completely anti-symmetric in exchanging the scalars. To obtain a Bose-symmetric amplitude, a simple possibility is to multiply this amplitude with the anti-symmetric structure constant of a Lie algebra with an adjoint index for each of the scalars, and to interpret the spin one particle as a photon. 

To study unitarity, it suffices to check the $s$-channel singularity. If $B$ appears as part of the residue of a pole in the s-channel, then
\begin{equation}
\lim_{s\rightarrow 0} B =  u \, ( \, \xi^R_1 \cdot p_2)
\end{equation}
which is a Mandelstam invariant times the product of two tree level three point amplitudes and an exchanged scalar. This is consistent with unitarity if and only if 
\begin{equation}
A \stackrel{?}{\propto} \frac{1}{s\,t } B \textrm{  or  }   A \stackrel{?}{\propto} \frac{1}{s\,u} B
\end{equation}
since in these cases the residue at the $s$-channel pole satisfies unitarity. For more concrete expressions, more information on the internal symmetry group structure is required.

The computation of the P-matrix in this case is straightforward as it is one by one, 
\begin{equation}\label{eq:singlguonthreescalars}
\sum_{\textrm{helicities}} B^2 = - \,s\, t \, u
\end{equation}
In this particular case the number of space-time dimensions, $D$, does not enter into this expression. This will turn out to be an exception: in general the P-matrix does depend on the dimension.

\subsection{One graviton and three scalars}
The basis of tensor structures in the convention introduced above reads
\begin{equation}
T_i = \{ \xi_1^L \cdot \xi_1^R, (\xi^L_1 \cdot p_2) (\xi^R_1 \cdot p_2), \ (\xi^L_1 \cdot p_2) (\xi^R_1 \cdot p_3), (\xi^L_1 \cdot p_3) (\xi^R_1 \cdot p_2), (\xi^L_1 \cdot p_3) (\xi^R_1 \cdot p_3) \}
\end{equation}
and contains five elements. At this point, the question is which field content one is interested in. For gravitons, one can first focus on a left-right symmetric ansatz for all graviton legs. In this particular case this reads
\begin{equation}
 \{ \xi_1^L \cdot \xi_1^R, (\xi^L_1 \cdot p_2) (\xi^R_1 \cdot p_2), (\xi^L_1 \cdot p_3) (\xi^R_1 \cdot p_3),(\xi^L_1 \cdot p_2) (\xi^R_1 \cdot p_3)+  (\xi^L_1 \cdot p_3) (\xi^R_1 \cdot p_2) \}
\end{equation}
which contains only four independent elements. Although this is a very minor difference in this simple case, in more complicated situations such as five gravitons this restriction to left-right symmetric states enables otherwise prohibitively difficult computations. In the case of one graviton, three scalars there turns out to be no difference in the result taking either Ansatz. The result of solving the on-shell gauge invariance constraints yields for a judicious normalisation choice a single solution
\begin{equation}
B =  (u \, \xi^L_1 \cdot p_2 - s \, \xi^L_1 \cdot p_3) (u \, \xi^R_1 \cdot p_2 - s \, \xi^R_1 \cdot p_3)
\end{equation}
Note that this is simply the product of two gluon amplitudes, \eqref{eq:onegluon}. This form is left-right symmetric: the anti-symmetric tensor does not couple here. Moreover, the graviton amplitude is completely symmetric under exchange of the scalar legs and hence obeys Bose-symmetry. 

To study unitarity it suffices to check the $s$-channel singularity. If $B$ appears as the residue of a pole then 
\begin{equation}
\lim_{s\rightarrow 0} B =  u^2 (\, \xi^L_1 \cdot p_2 ) ( \, \xi^R_1 \cdot p_2)
\end{equation}
This is proportional to the product of a two-scalar, one graviton amplitude and a three scalar amplitude, multiplied by $u^2$. There is a unique and local Ansatz which is consistent with this $u^2$ in the numerator:
\begin{equation}
A^0  \propto \frac{1}{s t u} B
\end{equation}
By the Bose symmetry of the numerator and denominator and the computation in the $s$-channel, this is unitary in all channels. 
 
To compute the P-matrix for the case at hand is a matter of a simple computation,
\begin{equation}
\sum_{\textrm{helicities}} B^2 = (s\, t \, u)^2
\end{equation}
which is simply the square of the single gluon P-matrix in equation \eqref{eq:singlguonthreescalars}. This typically does not happen.

\subsection{Two gluons and two scalars}
Next consider the scattering of two gluons and two scalars (see also \cite{Glover:2003cm}). The basis of tensor structures in the convention introduced above reads
\begin{equation}
T_i = \{ \xi_1 \cdot \xi_2, (\xi_1 \cdot p_2) (\xi_2 \cdot p_1),  (\xi_1 \cdot p_3) (\xi_2 \cdot p_1),   (\xi_1 \cdot p_2) (\xi_2 \cdot p_3),  (\xi_1 \cdot p_3) (\xi_2 \cdot p_3) \}
\end{equation}
and contains $5$ elements. These elements have homogenous mass dimension individually, but do not (have to) have the same mass dimension overall. There are two equations derived from on-shell gauge invariance:
\begin{equation} \begin{array}{c}
 \{ p_1 \cdot \xi_2,  \frac{1}{2} s \,(\xi_2 \cdot p_1),  \frac{1}{2} u\, (\xi_2 \cdot p_1),   \frac{1}{2} s\, (\xi_2 \cdot p_3), \frac{1}{2} u \,(\xi_2 \cdot p_3) \}\cdot \vec{\beta}   = 0\\
 \{ p_2 \cdot \xi_1, \frac{1}{2} s \, (\xi_1 \cdot p_2),  \frac{1}{2} s  \,(\xi_1 \cdot p_3),  \frac{1}{2} t  \,(\xi_1 \cdot p_2), \frac{1}{2} t \, (\xi_1 \cdot p_3) \} \cdot \vec{\beta}  = 0
\end{array}\end{equation}
which lead to four equations after stripping off the independent tensor structures,
\begin{equation}\left( \begin{array}{ccccc}
2 &   s &   u&  0& 0 \\ 
 0& 0& 0&    s&  u \\
 2&  s&  0&   t& 0 \\
0& 0&   s&  0&  t  
\end{array}\right) \vec{\beta} = 0
\end{equation}
The mass dimensions of the tensor structures are reflected in those of the coefficients of this matrix. This system of equations is dependent leaving two independent solutions after row reduction. Starting with this case more non-trivial choices need to be made for writing down explicit solutions. The aim in this article for tensor structures will be for local, completely symmetric expressions, factoring out as much as feasible overall momentum factors. In the case at hand, these expressions can be chosen as
\begin{equation}\label{basis:2gluon2scalar-1}
B_1  =  2 \, (p_2 \cdot \xi_1)(p_1 \cdot \xi_2) - s (\xi_1 \cdot \xi_2)
\end{equation}
and
\begin{multline}\label{basis:2gluon2scalar-2}
B_2  = -2 \,t\, (p_3 \cdot \xi_1) (p_1 \cdot \xi_2) + 2 \,(s+t) \,(p_2 \cdot \xi_1) (p_3 \cdot \xi_2) + 2\, s \,(p_3 \cdot \xi_1)(p_3 \cdot \xi_2)  \\  - 
 ( s \,t    + t^2) \, (\xi_1 \cdot \xi_2)
\end{multline}
The first solution has mass dimension $2$, while the second has mass dimension $4$. Both elements are Bose-symmetric in the gluon and in the scalar particles, which in this simple case actually arose directly after row-reduction and didn't require further processing. Due to the conventions for solving momentum conservation and transversality, this symmetry is not manifest. Furthermore, the first solution simply corresponds to a $\phi^2 F^2$-type local operator evaluated on-shell. 

Suppose one is interested in minimally coupled scalar theory. In the approach here, this implies the assumption of two types of three point amplitudes: two scalars and a gluon, and three gluons. Perturbative unitarity then fixes the poles and their residues in all the channels. There is some interesting physics in this exercise: let us write down the most general three point amplitudes,
\begin{equation}
A(1_g,2_s,3_s) = T^{a_1}_{i_2,j_3} g_1 \xi_{1}^{\mu} (p_2 - p_3)_{\mu} \qquad A(1_g,2_g,3_g) = g_2 f^{a_1,a_2,a_3} A^{\textrm{co}}(1_g,2_g,3_g)
\end{equation}
with $A^{\textrm{co}}(1_g,2_g,3_g)$ the known `colour-ordered' Yang-Mills amplitude. First take the tensors $T$ and $f$ to be generic tensors. One can derive residues at the pole as
\begin{equation}
\lim_{s\rightarrow 0} s A(1_g,2_g,3_s,4_s) = g_2 g_1 T^e_{i_3 i_4} f_e{}^{a_1,a_2}{ [\, 2(p_2\cdot \xi_1)(p_3 \cdot \xi_2) - 2(p_1\cdot \xi_2)(p_3 \cdot \xi_1) - t\,(\xi_1\cdot \xi_2)\, ]}
\end{equation}
in the $s$-channel, 
\begin{equation}
\lim_{t\rightarrow 0} t A(1_g,2_g,3_s,4_s) = 4 g_2^2 T^{a_2}_{i_3 j} T^{a_1}_{j i_4} (p_4 \cdot \xi_1) (p_3 \cdot \xi_2) 
\end{equation}
in the $t$-channel and 
\begin{equation}
\lim_{u\rightarrow 0} u A(1_g,2_g,3_s,4_s) = 4 g_2^2 T^{a_1}_{i_3 j} T^{a_2}_{j i_4} (p_3 \cdot \xi_1) (p_4 \cdot \xi_2) 
\end{equation}
in the $u$-channel respectively by perturbative unitarity. The two solutions to the on-shell constraints can also be studied in these limits. For the $s$-channel one obtains
\begin{align}
\lim_{s\rightarrow 0}  B_1  & =  2 (p_2 \cdot \xi_1)(p_1 \cdot \xi_2) \\
\lim_{s\rightarrow 0}  B_2 &  = -2 \,t\, (p_3 \cdot \xi_1) (p_1 \cdot \xi_2) + 2\, t \,(p_2 \cdot \xi_1)(p_3 \cdot \xi_2)
 -  t^2 \, (\xi_1 \cdot \xi_2)
\end{align}
Note that these remain independent vectors in the limit. Furthermore, the second is proportional to the unitarity-derived residue, with proportionality factor $t /(T^e_{i_3 i_4} f_e{}^{a_1,a_2}) $. In the $t$ channel, 
\begin{align}
\lim_{t\rightarrow 0}  B_1  & =  2 (p_2 \cdot \xi_1)(p_1 \cdot \xi_2) - s (\xi_1 \cdot \xi_2) \\
\lim_{t\rightarrow 0}  B_2 &  =  2 \,s \,(p_4 \cdot \xi_1) (p_3 \cdot \xi_2) 
\end{align}
and the Bose-related result for the $u$-channel. The most general Ansatz consistent with power counting, gauge invariance, Poincar\'e invariance and locality is
\begin{equation}
A(1_g,2_g,3_s,4_s) = A_s g_2 g_1 T^e_{i_3 i_4} f_e{}^{a_1,a_2}  + A_t g_2^2 T^{a_2}_{i_3 j} T^{a_1}_{j i_4} + A_u g_2^2 T^{a_1}_{i_3 j} T^{a_2}_{j i_4}
\end{equation}
with $A_{s}$, $A_{t}$ and $A_{u}$ functions of $s$ and $t$. By unitarity, these functions can have poles in the $s$, $t$ and $u$ channels. It is tempting to write an Ansatz
\begin{equation}
A_s \stackrel{?}{=} (\frac{\alpha_{1,s}}{s} +  \frac{\alpha_{2,s}}{t}  +  \frac{\alpha_{3,s}}{u} )B_1 +( \frac{\beta_{1,s}}{s t} +  \frac{\beta_{2,s}}{t u}  +  \frac{\beta_{3,s}}{s u} )B_2
\end{equation}
and similar expressions for $A_t$ and $A_u$. However, this Ansatz is over-complete, since
\begin{equation}
\frac{1}{s t} = -\frac{1}{u s} - \frac{1}{u t} 
\end{equation}
by momentum conservation. Hence
\begin{equation}
A_s \stackrel{?}{=} (\frac{\alpha_{1,s}}{s} +  \frac{\alpha_{2,s}}{t}  +  \frac{\alpha_{3,s}}{u} )B_1 +( \frac{\beta_{1,s}}{s t} +  \frac{\beta_{2,s}}{t u} )B_2
\end{equation}
is a better Ansatz. The $\alpha$ and $\beta$ parameters are now pure coupling constants, independent of the Mandelstam invariants, and can be fixed by considering limits without loss of generality. From the three channels one obtains three constraints. The $B_1$ coefficients $\alpha$ are zero. The $\beta$ coefficients are fixed after considering two limits, say the $s$ and $t$ channel ones. The remaining channel constraint is then impossible to satisfy, \emph{if there is no further relation between the $f^{abc}$ and $T^a_{ij}$ coefficients}. The way out is to have such a relation
\begin{equation}
g_2 [T^{a}, T^{b}] = \frac{1}{2} g_1 f^{ab}_c T^c
\end{equation}
which is nothing but the definition of a Lie algebra, up to a choice of normalisation! The upshot therefore is that massless spin one particles can only couple to scalars if their coupling constants involve a Lie algebra structure. The physical compact realisation of the Lie algebra can be derived as a consequence of a different version of unitarity: that of reality of the three particle amplitude.

The P-matrix in this case is a two by two matrix,
\begin{equation}
P= \left(\begin{array}{cc} (-2 + D) \, s^2 &  (-4 + D) \, s\, t\, (s + t) \\
 (-4 + D) \, s\, t\, (s + t)&    (-2 + D) t^2 \, (s + t)^2   \end{array} \right)
\end{equation}
This matrix is symmetric by construction. Note that its entries are generically symmetric under interchange of particles $1,2$ and $3,4$ by Bose symmetry. Its determinant reads
\begin{equation}
\det(P)= 4 (-3 + D) s^2\, t^2 \, u^2
\end{equation}
The root at $D=3$ of the determinant is physically the observation that in three dimensions a vector boson is Hodge-dual to a scalar: scalar amplitudes are all proportional to each other, so there is really only one independent solution to the on-shell constraints. Thought about differently, this results shows that by considering only on-shell kinematics we have arrived at three dimensional Hodge-duality.

\subsection{Two gravitons and two scalars}
Consider the scattering of two gravitons with two scalars. The same computation as the two gluon case treated above will be followed, emphasising the differences. The basis of tensor structures is now taken to consist of left-right symmetric products of polarisations vectors for each leg, e.g.  
\begin{equation}
(\xi^L_1 \cdot \xi^L_2)(\xi^R_1 \cdot \xi^R_2) + \textrm{permutations}
\end{equation}
is one of the $T_i$ elements, and the permutations range over all left-right swaps of polarisation vectors. This is a restricted Ansatz which contains both dilatons and gravitons for the external legs. Below we comment on projecting out these states. There are $14$ elements in the $T_i$ basis using the conventions introduced above. Deriving the on-shell gauge invariance equations gives $12$ equations per external graviton leg. Row-reducing the set of equations gives $3$ solutions. Before turning to basis elements, consider the P-matrix constructed out of the raw output of Mathematica. The determinant of this three by three matrix is
\begin{equation}
\det(P) \propto (D-3 )^2 (D-2)
\end{equation}
up to functions of Mandelstam invariants. This determinant is relatively basis-independent. The power of $D-3$ indicates that we can treat the symmetrised external state as a scalar in three dimensions: there is only one independent amplitude in this number of dimensions. Since we did not subtract off the trace from the states, the conclusion must be that only the trace part of the graviton couples non-trivially. This of course is the well-known observation that in three flat dimensions gravity does not have any local propagating degrees of freedom. It is interesting that from the point of view developed here, the triviality of three dimensional gravity is directly related to hodge duality for spin one particles. The appearance of the single two-dimensional relation is intriguing, but we have no intrinsic explanation. 

Although in this case explicit solutions to the on-shell constraints are easy to obtain directly, we will follow a slightly different route to obtain a nice form of the basis in this case. The simple but crucial observation is that a special class of solutions to the on-shell constraints for gravitons can be obtained by multiplying solutions of the on-shell constraints for gluons. Since these were obtained in a Bose-symmetric form above, the product of these gluon factors is automatically Bose-symmetric. They are however not left-right symmetric yet. At the very least, the solutions should be invariant under exchanging the full set of `left' and `right' polarisations, i.e. the starting point can be taken to be:
\begin{align}
\tilde{B}_1 &= (B_1)_L(B_1)_R \\
\tilde{B}_2 &= (B_2)_L(B_2)_R \\
\tilde{B}_3 &= (B_2)_L(B_1)_R + (B_1)_L(B_2)_R
\end{align} 
which contains three elements. These three elements span all solutions to the on-shell constraints for the two-graviton amplitude. To prove this statement two steps are necessary. First, these solutions need to be symmetrised over  left-right swaps of the individual legs. Then, it needs to be verified that the resulting three solutions are linearly independent as vectors in the vector space spanned by the relevant 14 $T_i$ tensor structures. In this particularly simple case, this turns out to be the case and the above is indeed a valid basis of solutions to the on-shell constraints after left-right symmetrisation.

Let us construct minimally coupled gravity amplitudes out of the above Ansatz. Without computation and based on the gluon experience gained above, it can be guessed that only $\tilde{B}_2$ will lead to a consistent amplitude. By dimensional analysis and Bose symmetry alone, this must be
\begin{equation}
M(\phi,\phi, G^{\textrm{sym}}, G^{\textrm{sym}}) = \frac{1}{s t u} \tilde{B}_2
\end{equation} 
Note that perturbative unitarity follows directly from the gluon analysis: this is seen most clearly by considering the minimally coupled scattering amplitudes without left-right symmetrisation,
\begin{equation}
M(\phi,\phi, G,G) = \frac{1}{s t u}  (B_2)_L (B_2)_R = s \left( \frac{B_2}{ s t}\right)_L  \left(\frac{B_2}{ s u}\right)_R  
\end{equation} 
which also cleanly exposes the appropriate double copy type relation.

\subsection{Three gluons and a scalar}
The next case contains three gluons and a single scalar particle. The basis of allowed tensor structures in our conventions contains fourteen elements. Solving the on-shell gauge invariance constraint then yields $4$ solutions. New in this case is the fact that the solutions do not have any obvious symmetry properties beyond homogeneity in mass dimension as output of solving the on-shell constraints. To construct a basis some choices must be made. First, make the solutions local by multiplying out any denominators. Then, one can aim to find solutions with prescribed symmetry by brute force search, factoring out as much as possible overall multiplicative factors of momentum. For this one can sum over the relevant orbit of the permutation group, permuting polarisations and momenta simultaneously. Given a scattering form $a$, a symmetric form is obtained as
\begin{equation}\label{eq:symmmastereq}
a_{\textrm{sym}}\, = \sum_{permutations} f(s,t) a
\end{equation}
for every choice of $f$. In this case we have four solutions and we should be looking for a basis of the same size. Using $f=1$ in the above formula for instance for the raw output of Mathematica in the case at hand gives zero independent elements: they all vanish in this sum. Scanning monomial choices for $f$ and checking linear independence then gives a result with four independent vectors for $f = s^4 t$. The resulting expressions now have fairly high mass dimension,
\begin{equation}
[\textrm{mass} ] = \{13,15,15,15\}
\end{equation}
However, simplifications are possible, using properties of the completely symmetric polynomials. For four particles (see e.g. \cite{Boels:2013jua} for higher points) there are two almost canonical basis polynomials
\begin{equation}\label{eq:complsymbas}\begin{array}{rl}
\sigma_1 & = s^2 + t^2 + u^2 \\ 
\sigma_2 & = s\,t\,u\end{array}
\end{equation}
which are manifestly invariant under any exchange of the four external legs. Every homogenous polynomial of homogeneity $h$ has a unique expansion in terms of all products of $\sigma_1$ and $\sigma_2$ with homogeneity $h$. 

To find simplifications of a given set of completely symmetric basis elements amounts to the question if there are linear combinations with coefficients which are homogeneous functions of completely symmetric polynomials such that the right hand side is proportional to a sum over completely symmetric polynomials times local basis elements. As a subset of this question, one can study the case where the right hand side is proportional to a specific completely symmetric polynomial,
\begin{equation}\label{eq:findsimpler}
\sum_i c_i (\sigma_1, \sigma_2) \tilde{B}_i \propto g_{\textrm{symmetric}}
\end{equation}
Up to mass dimension $12$, the only polynomials which can appear on the right hand side are integer powers of either $\sigma_1$ or $\sigma_2$. At mass dim $12$ there are two such polynomials: $\sigma_1^3$ and $\sigma_2^2$. This yields a practical method to search for coefficients $c_i$ for factoring out low degree polynomials. First fix an overall mass dimension for the equation sought for. Then expand $c_i$ as completely symmetric polynomials into the polynomial basis. 
\begin{equation}
c_{i}  = \sum_j c_{ij} g_{j} (\sigma_1,\sigma_2)
\end{equation}
This yields a large Ansatz with unknown numeric coefficients $c_{ij}$. Now evaluate this Ansatz on a root of either $\sigma_1$ or $\sigma_2$. Then the search is for a linear relation between a set of fixed, numerical vectors. This can be solved using the generic technique for relation-finding  using random integers described above. Having found the coefficients $c_{ij}$ one can now plug these into equation \eqref{eq:findsimpler} and find a right-hand side which is a simpler basis element times an overall factor. One can now construct a new basis by exchanging one of the old basis elements by the new one. In this way, one can systematically lower the basis dimensions. In this particular case one obtains a symmetric basis with minimal mass dimension
\begin{equation}
[\textrm{mass} ] = \{7, 9, 9, 11\} \qquad \textrm{symmetric basis}
\end{equation}
One can in this case actually obtain smaller minimal mass dimensions when one constructs a basis for which the elements are completely anti-symmetric under exchange of each pair of the gluon elements.  
\begin{equation}
[\textrm{mass} ] = \{3, 5, 7, 9\} \qquad \textrm{anti- symmetric basis}
\end{equation}
The construction is similar to the one just outlined for symmetric basis choices, but one inserts the anti-symmetric symbol into the permutation sum.  

One can in this case also compute the P-matrix for each choice of basis element. The determinant of the four by four matrix is
\begin{equation}
\det(P) \propto (D-3 )^3
\end{equation}
as should be expected by now by Hodge-duality: there are indeed three relations between the four amplitudes in three dimensions.

\subsection{Three gravitons and a scalar}
In this case the starting point is the left-right symmetric Ansatz, which contains $85$ elements. Solving the on-shell gauge invariance constraint yields $11$ independent solutions. This is a very interesting number, as by naive squaring of the three gluon amplitudes only $4*5 /2 = 10$ overall-left-right symmetric solutions are obtained. Hence, this counting demonstrates that there can be solutions to the on-shell constraints for graviton amplitudes which cannot be constructed by taking sums over products of gluon amplitudes. The set of left-right symmetrised products of gluon amplitudes can be checked in this case to be linearly independent. 

To construct Bose-symmetric amplitudes, the solutions to the on-shell constraints can be symmetrised on the graviton legs using the technique above. After a considerably harder computation especially when minimising the mass-dimension by factoring out overall completely symmetric polynomials, one arrives at a semi-minimal, Bose-symmetric, left-right symmetric local basis of eleven elements with mass dimensions
\begin{equation}
[\textrm{mass} ] = \{ 6, 6, 8, 8, 10, 10, 10, 12, 12, 12, 14    \} \qquad \textrm{symmetric basis}
\end{equation}
Next in this computation is the projection of the obtained gravitational solutions onto the sums of products of gluon amplitudes. This can be done by vector algebra. The problem is formulated as follows: given two sets of vectors $S_1 = \{\vec{a}_1, \vec{a}_2, \ldots, \vec{a}_m \}$ and $S_2 =  \{\vec{b}_1, \vec{b}_2, \ldots , \vec{b}_k\}$ which are both linearly independent independently, find all linear combinations of one set in terms of the other. This amount to constructing the matrix
\begin{equation}
\{\vec{a}_1, \vec{a}_2, \ldots, \vec{a}_m,\vec{b}_1, \vec{b}_2, \ldots , \vec{b}_k\} 
\end{equation}
and using Mathematica to compute its kernel. This kernel is exactly the sought-for set of relations. Mathematica employs `inverted' rowreduction in its `NullSpace' command: the output is in lower triangular form, for a horizontally flipped version of the usual lower triangular form. This triangularity makes the output easily readable. Furthermore, it also exposes that vector in the set $S_2$ which does \emph{not} have a linear relation to the vectors in set $S_1$. Up to additions of sums over products of gluon amplitudes, this is an exceptional gravitational amplitude which cannot be written as sums over products of gluon amplitudes.  In this particular case it has minimal mass dimension $6$. For the amplitudes in question the `Nullspace' command yields a set of vectors which are natural functions of $\sigma_1$ and $\sigma_2$  by the symmetry properties of all amplitudes involved.

Factoring out overall completely symmetric polynomials does not guarantee a \emph{minimal} basis. Indeed, there can be linear combinations of basis vectors which yields a linear combination of lower dimension local building blocks which are not found by the above technique. We have observed this in examples. On the other hand, the local, symmetric products of gluon amplitudes would yield mass dimension
\begin{equation}
[\textrm{mass} ] = \{ 6, 8, 10, 10, 12, 12, 14, 14, 16, 18    \} \qquad \textrm{gluon products}
\end{equation}
which is generically larger than the result above. The relation between the two basis constructions has just been discussed. Although we are confident the factorisation problem may be solved uniquely, this leads beyond the scope of this article. 

Using integer values for the momenta one can obtain the determinant of the P-matrix in this case. The exact form of the basis used to derive this is unimportant if one is foremost interested in dimension dependence. In this case the P-matrix computation is easiest in terms of the Mathematica output for solving the on-shell constraints. The dimension-dependent part of the determinant of the P-matrix reads
\begin{equation}
\det(P) \propto (-3 + D)^{10} (-2 + D)^6 (4 - 3 D + D^2)
\end{equation}
The three dimensional root is the triviality of three dimensional gravity as observed above. The last factor has rather intriguing imaginary roots, while the power of $D-2$ is certainly also remarkable. An explanation for these could be very interesting.

\section{Four gluons / Four gravitons: solution of on-shell constraints}\label{sec:fourpnoscal}

After the examples involving at least a scalar particle in the previous section the stage is set to move to four gluon and four graviton scattering amplitudes. Compared to the previous cases there will be some new features. Due to the computational complexity of intermediate steps, only salient details will be highlighted below. 

\subsection{Constructing a basis for four gluons}
This case was discussed briefly in \cite{Boels:2016xhc}, and recently also in more detail in \cite{Bern:2017tuc}. The latter also featured an explicit basis choice for a subspace of the solutions to the on-shell constraints. Generating four gluon tensor structures yields an ansatz of $43$ elements in our conventions. In the four gluon case at tree level, it is also possible to take a restricted ansatz by dropping those terms which do not feature a metric contraction between external gluon polarisation vectors, i.e. which only involve $\xi_i \cdot p_j$ type terms. Put positively, this restricted Ansatz would contain at least single metric contractions between external polarisation vectors, i.e. at least a single $\xi_i \cdot \xi_j$ type term. For this restricted Ansatz only a single solution is obtained after implementing on-shell gauge invariance. This solution is always proportional to the Yang-Mills four point amplitude. Widening the Ansatz to the full $43$ element case gives $10$ solutions from on-shell gauge invariance. The determinant of the P-matrix gives up to kinematic dependence
\begin{equation}\label{eq:fourgluondetP}
\det(P) \propto (-4 + D)^2 (-3 + D)^9 (-1 + D) \, .
\end{equation} 
The four gluon amplitude showcases the first appearance of two so-called evanescent terms in four dimensions. These are terms that are non-zero in general dimensions, but vanish in strictly four dimensional kinematics, that is, with polarisation vectors and momenta in a common four dimensional subspace. The precise evanescent terms can be computed as the kernel of the P-matrix when $D$ is set to four. One way to understand the evanescent terms is that four gluon amplitudes can have maximally $2^4 = 16$ different helicity amplitudes, distributed over $8$ parity even and $8$ parity odd structures\footnote{This argument was essentially also made in \cite{Kravchuk:2016qvl}. }. As will be argued below in section \ref{sec:oneloop},  there are actually only five independent four-gluon amplitudes through at least two loops.  
  
The first step in constructing the basis is to split the solutions into mass-dimension-homogeneous solutions. To obtain a local solution all solutions are multiplied by their overall poles, yielding a manifestly local expression. In the case of four gluons symmetry is more complicated (and interesting!) than the previously studied examples. Scanning for completely symmetric basis elements shows a surprising new feature: no matter how complicated (high mass dimension) a function $f$ is inserted in equation \eqref{eq:symmmastereq}, one always obtains maximally seven independent completely symmetric tensor structures. 

The crucial observation is the following: the only polynomial of Mandelstams anti-symmetric in the exchange of three legs is also anti-symmetric in the exchange of four legs. The latter polynomial is unique and proportional to
\begin{equation}\label{eq:complantisym}
\propto (s-u)(t-s)(u-s)
\end{equation}
Hence a tensor structure can constructed such that it is anti-symmetric in three of the legs 
\begin{equation}\label{eq:nonsymmmastereq}
a^1_{\textrm{non-sym}} = \sum_{\pi \in \textrm{permutations}}  g_{\{2,3,4\}}(\pi) \, f(s,t) \, a
\end{equation}
Here the upper index ``1'' means single out the element ``1'' and $g_{\{2,3,4\}}$ gives the signature of the permutation of the elements $2,3,4$ inside the permutation $\pi$. The resulting amplitude cannot be related to the completely symmetric solution to the on-shell constraints by multiplying with functions of Mandelstams. The simplest local solution to the on-shell constraints obtained this way is of mass dimension $8$. Since the procedure singles out a single gluon leg, one can wonder about the full set $a^i$. It turns out that there is one simple relation between these four tensor structures:
\begin{equation}
a^1- a^2 + a^3 - a^4 =  0 
\end{equation}
Hence, there are three independent mass dimension eight solutions of this particular symmetry type. Since there are ten solutions to the on-shell constraints in total this implies that there are indeed only seven completely symmetric tensor structures. Knowledge of this number makes the analysis of the symmetric tensor structures much simpler. The tensor structures $a^i$ will be referred to as `non-symmetric tensors'. 

After generating a maximal seven-dimensional set of completely symmetric tensor structures by summing as in equation \eqref{eq:symmmastereq}, one can again factor out overall powers of the completely symmetric polynomials using the technique introduced above. The result is a set of mass dimension
\begin{equation}
[\textrm{mass} ] = \{ 4, 4, 6, 6, 6, 8, 8  \} \qquad \textrm{symmetric basis}
\end{equation}
where the first element is taken to be 
\begin{equation}
B_1 = A^{\textrm{YM}}(1,2,3,4) \, s\,  t
\end{equation}
where $A^{\textrm{YM}}(1,2,3,4) $ is the colour-ordered Yang-Mills tree amplitude for unit coupling constant. Taken together with the non-symmetric three basis elements above, these seven structures form a basis for all gluon four point amplitudes. For definiteness, one could define
\begin{equation}
\tilde{B}_8 = a^1 \qquad \tilde{B}_9 = a^2  \qquad \tilde{B}_{10} = a^3 
\end{equation}
Alternatively, one could take three tensor structures all anti-symmetric in the same three legs. In that case, the minimal dimensions are $8,10,12$. The latter is the choice we make here, in part for legacy reasons.

A key observation is that the P-matrix in the four gluon case in the basis constructed is block diagonal. The driver here is the equation
\begin{equation}
\sum_{\textrm{helicities}}B^{\textrm{sym}}B^{\textrm{non-sym}} = 0
\end{equation}
which can be checked easily. Hence, the P-matrix in the basis constructed so far splits into two blocks: a seven by seven block for the symmetric tensors (denoted $P_7$) and a three by three block (denoted $P_3$) for the non-symmetric tensors, e.g.
\begin{equation}
P = \left(\begin{array}{cc} P_7 & 0 \\ 0 &P_3 \end{array} \right)
\end{equation}
The sub-matrices have determinants
\begin{align}
\det(P_7) & \propto (-4 + D)^2 (-3 + D)^6 (-1 + D)\\
\det(P_3) & \propto (-3+D)^3
\end{align}
This shows the tensor structures in $P_3$ decouple completely in three dimensions. Furthermore, there are two relations between the symmetric tensor structures contributing to $P_7$ in four dimensions. These can also be found explicitly, by computing
\begin{equation}
\textrm{Ker} \left[ P_7(D=4) \right]
\end{equation}
These give two independent linear combinations of the tensor structures which vanish in four dimensional kinematics, which will be taken to be $B_6$ and $B_7$. The basis constructed up to this point then splits into three parts
with five, two and three elements respectively, with mass dimensions
\begin{equation}
[\textrm{mass} ] = \{ 4, 4, 6, 6, 8 | 8, 10 | 8, 10, 12    \} 
\end{equation}
The explicit expressions for this basis are included in the arXiv submission. Note that although external states can always be chosen to be in four dimensions by an appropriate scheme choice (e.g. \cite{Bern:1991aq}), such luxury does not exist for the internal states running in the loops in dimensional regularisation. This is well known from studies of the dimensional reduction renormalisation scheme \cite{vanDamme:1984ig} \cite{Jack:1993ws}. 

The basis thus constructed is independent for generic momenta. However, at special loci relations between the basis elements may arise. This is closely related to the factoring out of completely symmetric polynomials out of a linear combination of tensor structures. A particular case consists of relations which contain the roots of the completely anti-symmetric polynomial \eqref{eq:complantisym}:
\begin{equation}
t\rightarrow - 2 s  \qquad s \rightarrow t \qquad s\rightarrow - 2 t 
\end{equation}
It turns out, one can find two relations which involve this polynomial between the basis vectors. Introducing these yields two new basis elements to replace two old ones. The new elements are such that when divided by $s \, t$ or multiplied by $u = -s-t$ the tensor structure is completely antisymmetric. 

\subsubsection{Unitarity and locality}
As in the above examples that involve at least a single scalar, also in the four gluon case one can construct explicit scattering amplitudes consistent with tree level unitarity. In particular, one can aim to introduce poles and study minimal coupling. The dimensionality of the local basis in the limit where one of the Mandelstams vanishes, such as $s \rightarrow 0$, does not change. This can be seen as follows: suppose a local relation arises in the limit  $s \rightarrow 0$. Then there exists polynomial coefficients such that
\begin{equation}
\lim_{s \rightarrow 0} \sum_i \alpha_i(s,t) B_i = 0
\end{equation}
which give before the limit
\begin{equation}
\sum_i \alpha_i(s,t) B_i = s \left( \textrm{local} \right)
\end{equation}
Since by construction the basis elements $B_i$ are independent one can use the expression on the right hand side as a new basis element, of smaller mass dimension. Hence, there is a minimal basis of local tensor structures. Since there are two independent three point amplitudes with three gluons this implies that there are three four-point tensor structures which can appear with an explicit simple pole. Since the three point amplitudes are generated by known actions, it is not hard to guess which amplitudes these must be. The usual Yang-Mills amplitude is particularly easy to identify since it is proportional to the one tensor structure which has at least one metric contraction between gluon polarisation vectors. Repeating the argument above for general coupling constants cleanly shows the colour-Jacobi relation as a necessary consistency condition.

\subsection{Constructing a basis for four gravitons}

The Ansatz for solving the on-shell constraints for four symmetric but not traceless external states in the graviton case contains $633$ elements. Restricting to Ansatz terms which have at least two metric contractions between the external polarisation vectors gives exactly one solution - this solution is proportional to the Einstein-Hilbert amplitude. Since the product of two four gluon Yang-Mills amplitudes also solves the on-shell constraints for gravitons, there must be a linear relation between the Einstein-Hilbert four graviton amplitude and this product: this is the physical origin of the KLT relation at four points \cite{Boels:2016xhc}. Considering terms with at least one metric contraction or the most general case of no metric contractions yields $11$ and $56$ solutions respectively. These solutions are not automatically Bose-symmetric. 

To find a reasonably good basis, first consider the double copy solutions to the on-shell constraints. There are naively
\begin{equation}
(7 \times 3 = 21)+ (7 \times 8 / 2 = 28) + (3 \times 4 /2 = 6) = 55 
\end{equation}
combinations which can be made out of gluon amplitudes that are symmetric under left-right exchange. Not all of these are linearly independent: from the $3 \times 4/2$ exactly three combinations are expressible in terms of the ``$7 \times 8 /2 = 28$'' set. Hence, there are $(56-52 = )$ four solutions to the on-shell constraints that cannot be written as gluon products. To find these, one expresses the $56$ explicit solutions in terms of the $52$ gluon products using matrix algebra. This also points out those solutions of the set of $56$ which cannot be written as squares of gluons. 

\begin{figure}\centering
\includegraphics[scale=0.5]{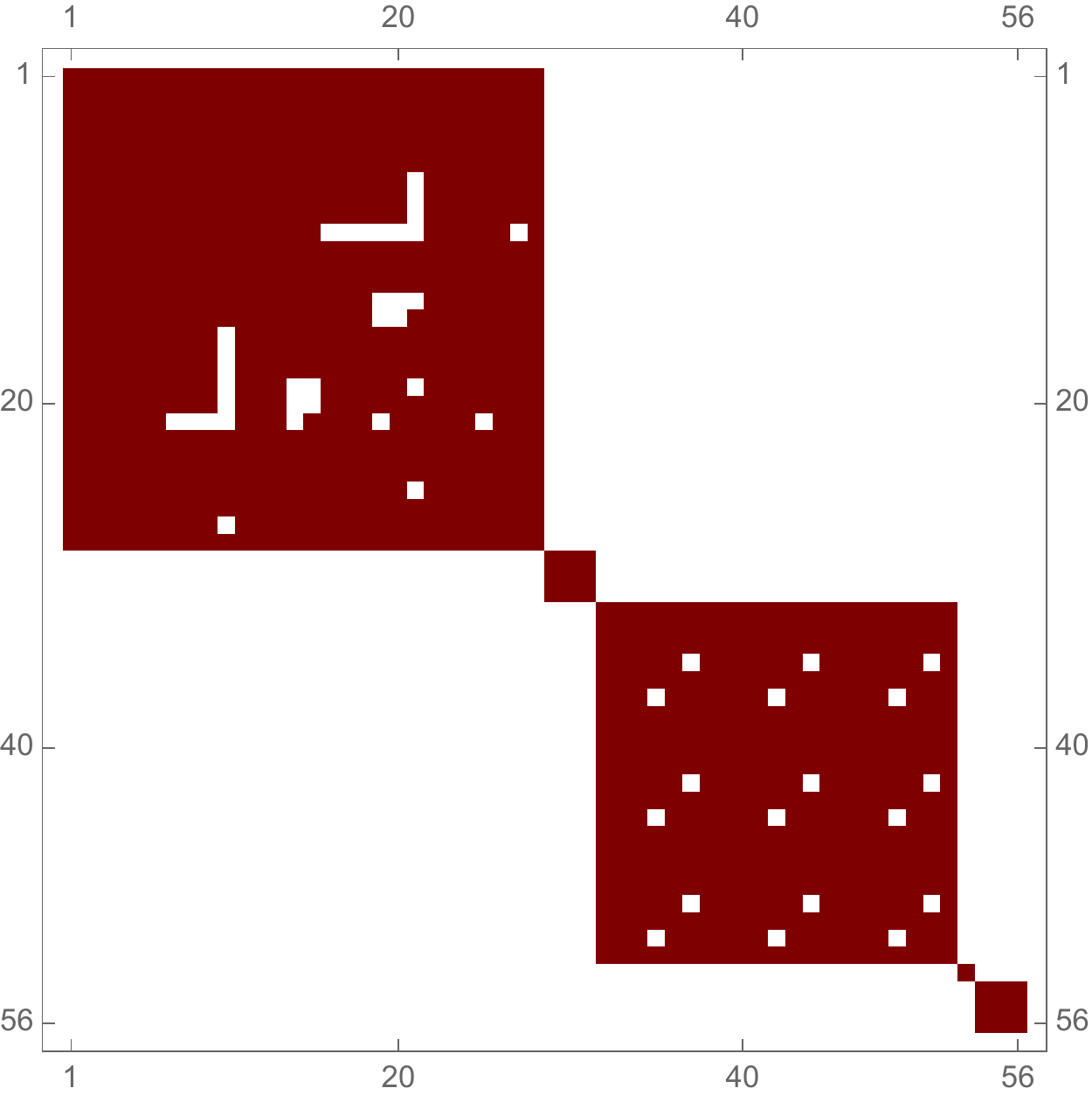}
\caption{Structure of the P-matrix for four gravitons, output of Mathematica command `MatrixPlot'. White indicates zero, burgundy non-zero entries. An explanation for the ``game of life'' pattern of zeroes within the blocks would be interesting.}
\label{fig:Pmatrixplot}
\end{figure}

Summarising we can construct a basis as follows: take the $28$, add the $3$ linearly independent elements out of the set of $6$, add the $21$ mixed elements and finally add the four non-double copy matrix elements. One can cross-check that these are linearly independent. What is even more important is that the P-matrix for this choice can, with some additional effort, be made block-diagonal with $28 \times 3 \times 21 \times 1 \times 3$ blocks, schematically
\begin{equation}
P = \left(\begin{array}{ccccc} P_{28} & 0 &0 & 0& 0 \\ 0 & P_3 & 0 & 0& 0 \\ 0 & 0 & P_{21} & 0& 0 \\ 0 & 0 & 0 & P_{1} & 0 \\ 0 & 0 & 0 &0 & P'_3\end{array} \right)
\end{equation}
See also the figure \ref{fig:Pmatrixplot}. The last four elements of the chosen basis are of the non-gluon square type. This block-diagonal shape facilitates inverting the P-matrix enormously. 

The origin of the elements allows one to infer some basic properties. The first $28$ elements are automatically Bose symmetric. After explicit symmetrisations to high order, it appears only one indepedentn amplitude remains. This involves the four non-double copy elements only. In the chosen basis, this is the first element of the last four entries. Orthogonality was enforced to find the other three. The dimension dependent part of the determinant of the full P-matrix for the case at hand is
\begin{multline}
\det(P) \propto (D-4)^{19} (D-3)^{55} (D-2)^{24} (D-1)^{24} \left(D^2- 3 D-2\right)^2 \\  \left(D^2-3 D+4\right)^4 \left(D^4-5 D^3+8 D^2-8 D+16\right)
\end{multline}
It can be checked from the P-matrix also explicitly that in the four dimensional limit $19$ relations arise between the $56$ solutions to the on-shell constraints. Since the external states are symmetric, but not traceless, this number can be understood to an extent by counting. In four dimensions, there are two helicity states for the graviton and one for the dilaton. The graviton states are parity conjugates, while the dilaton is parity neutral. Hence, when counting the parity even scattering amplitudes, there are $8$ parity even four graviton amplitudes, $4 \times 8 /2 = 16$ parity even three graviton, single dilaton amplitudes, $(3 \times 4 /2)\times 4 /2 = 12$ two parity even two graviton - two dilaton amplitudes,  $4$ parity even single graviton amplitudes and a single four dilaton amplitude. This yields $41$ independent parity-even helicity states. Hence there are more helicity amplitudes than solutions to the on-shell constraints in this particular case. It would be interesting to understand this in more detail, especially as we present evidence below for additional, general relations between helicity amplitudes involving gravitons at loop level. 

The explicit form of the four dimensional relations can be obtained from the kernel of the P-matrix at the point $D=4$. If only a set of independent elements is required, one can compute the kernel numerically and identify the dependent amplitudes. The appearance of non-integer and even complex-valued roots for the dimension dependence of the determinant is certainly striking - this remains to be understood.

\subsection{Application: Open and closed string four point amplitudes}
A point to be stressed time and again is the fact that any scattering amplitude is expressible in terms of a basis of kinematic factors as obtained above explicitly in the four particle case. Consider for instance the four point, tree level gluon amplitude in bosonic string theory. This can be computed using textbook worldsheet methods. An explicit answer can be found in the literature, for instance in the original KLT paper \cite{Kawai:1985xq}. In the conventions of this article, this amplitude can be written in the following computer-friendly form
\begin{multline}
A_4 = G(\alpha' s,\alpha' t)  (\alpha' s - 2 ) (\alpha' t - 2 ) (\alpha' u - 2 ) \left[ \frac{s}{2}  (\xi_1 \cdot p_3 \xi_2 \cdot p_3 (\xi_3 \cdot p_1 \xi_4 \cdot p_1 + \xi_3 \cdot p_2 \xi_4 \cdot p_2)   )  \right. \\
\left.\frac{s \, t}{ 2 (\alpha' u - 2 )} ( \xi_1 \cdot \xi_3 - \xi_1\cdot p_3 \xi_3 \cdot p_1)  ( \xi_2 \cdot \xi_4 - \xi_2\cdot p_4 \xi_4 \cdot p_2)  + \textrm{permutations} \right]
\end{multline}
where $G$ is essentially the ${}_2 F_1$ hypergeometric function,
\begin{equation}
G(-\alpha' s,-\alpha' t) = \frac{\Gamma(\alpha' s) \Gamma(\alpha' t)}{\Gamma(- \alpha' s- \alpha' t +1)}
\end{equation}
and the sum over permutations includes all simultaneous exchanges of momenta and polarisation vectors of the four external legs. Projecting onto the basis using either linear algebra or P-matrix is possible. In terms of the mostly symmetric basis found above a vector is obtained for $\alpha'=1$,
\begin{multline}
A_4 \propto \left\{\frac{1}{4} (\sigma_1 -2) \sigma_1+\sigma_2,-\frac{3}{4} (\sigma_1-2) \sigma_1-3 \sigma_2,-2 \sigma_2, \right. \\  \left. -\sigma_1-2 \sigma_2+2,2 (\sigma_1+2 \sigma_2-2),\sigma_1-2,1-\frac{\sigma_1}{2},0,0,0\right\}
\end{multline}
Since the part of the amplitude which contains the polarisation dependence is completely Bose-symmetric, the non-symmetric coefficients listed last are zero.

The closed string amplitude can be obtained from the open string amplitude by the KLT relation,
\begin{equation}
A_{cl} = A_{o, L}(1,2,3,4) A_{o, R}(1,3,2,4) \sin(\alpha' t) 
\end{equation}
This relation can be understood as the only possible solution to the problem of creating an amplitude with non-overlapping single particle poles out of the product of two open string amplitudes. In effect, the Veneziano-like ratio of Gamma functions is simply replaced by the Virasoro-Shapiro-like closed string structure. The fact that a relation between gluon and graviton scattering exists is covered by on-shell gauge invariance. The exact form of the relation is determined to a large extent by unitarity and locality, e.g. having only local simple poles. It would be interesting to investigate the exact extent.

For the basis for gravitational amplitudes chosen here, only the $28$ first elements get non-zero coefficients. This is inherited from the vanishing of the non-symmetric gluon coefficients, as well as the fact that this amplitude is obtained from a double copy relation: the non-squaring coefficients vanish for the closed string tree amplitude. Apart from inherent interest, the string theory amplitude is useful to perform consistency checks on computer code.

\section{One-loop amplitudes}
\label{sec:oneloop}

Given a set of tree level amplitudes, it is a fundamental problem to construct the perturbative quantum corrections contained in the loop amplitudes from this set. That these corrections have to be there is a consequence of perturbative unitarity: discontinuities across branch cuts are given by products of lower-loop amplitudes, summed over internal helicities and integrated over integral on-shell momenta. Since the latter is neither zero nor integrates to zero in general, there must be non-vanishing loop amplitudes, generically to all loop orders \footnote{A somewhat contrived counterexample to the `all orders' part of this statement is the self-dual Yang-Mills theory: this has only a single tree amplitude, one infinite series of one-loop amplitudes but nothing beyond.}. In this section it will be shown how solving the on-shell constraints can be used to compute loop amplitudes using essentially just unitarity and locality. The general strategy and its viability are shown in this section in one loop computations. $D$ dimensional unitarity cuts will be taken using Cutkosky's rules \cite{Cutkosky:1960sp}.

\begin{figure}\centering
\includegraphics[scale=0.5]{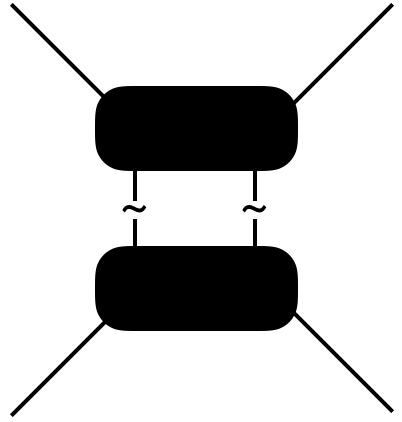}
\caption{discontinuity across branch cut for four point amplitude.}
\label{fig:oneloop4ptcut}
\end{figure}

For clarity in this section the spin is increased gradually, starting at amplitudes with four external scalars. Several details of the calculation will only be highlighted once, and the section is meant to be read largely consecutively. The section on four gluons is intentionally reasonably self-contained for expert (or impatient) readers, who are directed there.

\subsection{Four scalars}

\subsubsection{the $\phi^3$ theory}
Consider $\phi^3$-theory, with Lagrangian 
\begin{equation}
\mathcal{L} = \frac{1}{2} (\partial_{\mu }\phi)(\partial^{\mu }\phi) - \frac{\lambda}{3!} \phi^3
\end{equation}
Although this Lagrangian is not necessary, it serves as a useful guide and cross-check for the computation below. The three point scattering amplitude is
\begin{equation}
A^0_{3}  = \lambda 
\end{equation}
The momentum conserving delta function which is routinely suppressed in amplitude expressions induces a coupling constant which is of mass dimension $1+2 \epsilon$ in $D=4 - 2 \epsilon$, assuming appropriately normalised fields.  The four point tree level scattering amplitude is
\begin{equation}
A^0_{4}  = \lambda^2 \left(\frac{1}{s} + \frac{1}{t} + \frac{1}{u} \right)
\end{equation}
which can be derived from assuming tree level unitarity, a spectrum of massless particles and a single dimension-full coupling constant $\lambda$. At the one loop level, unitarity states that the discontinuity across a given branch cut is given in terms of a integral of over the un-observed on-shell particles in the product of two tree amplitudes, e.g.
\begin{equation}
\textrm{disc}_{s} A^1(s,t) = \int d\textrm{LIPS}_2 A^0(1,2,L_1, L_2)A^0(3,4,-L_1,- L_2)
\end{equation}
See the figure \ref{fig:oneloop4ptcut} for an illustration. There are two additional constraints for cuts in the  $t$ and $u$ channels. Momentum conservation can be solved on both sides of the cut by choosing appropriate Mandelstams,
\begin{equation}
s = (p_1+p_2)^2 \qquad t = (p_2+p_3)^2 \qquad t_L = (L_1 + p_2)^2 \qquad t_R  = (L_1-p_3)^2
\end{equation}
Putting in the tree level expressions gives:
\begin{equation}\label{eq:cutsphicubed}
\textrm{disc}_{s} A^1(s,t) = \lambda^4 \int d\textrm{LIPS}_2 \left( \frac{1}{s} + \frac{1}{t_L} - \frac{1}{t_L + s}\right) \left( \frac{1}{s} + \frac{1}{t_R} -\frac{1}{t_R + s}\right)
\end{equation}
These expressions also arise from Cutkosky's rules for cutting scalar integrals. For instance, for the bubble integral Cutkosky's rule gives
\begin{equation}
\textrm{disc}_{s}  \left[\int d^DL_1 \frac{1}{L_1^2 (L_1 + p_1 + p_2)^2}   \right]=   \int d\textrm{LIPS}_2 \left(1 \right)
\end{equation}
All cut terms in equation \eqref{eq:cutsphicubed} can unambiguously be assigned to Cutkosky-style cut scalar box, triangle and bubble integrals. The triangle and bubble integrals for four external massless particles are all single scale integrals, and can be computed straightforwardly. Since these are all proportional to $s$,$t$ and $u$ raised to an appropriate power, one can express all these integrals in a basis of, say, the bubble integrals which share this property. Define furthermore the box functions
\begin{equation}
\textrm{Box}(s,t) = \int d^D L \frac{1}{L^2 \,(L + p_1 + p_2)^2\, (L + p_2)^2\,  (L-p_3)^2}
\end{equation}
Note that the box functions each have a cut in two channels whose subscript appears among the labels given. A complete basis for all integrals that appear is formed by $\textrm{Box}(s,t) $, $\textrm{Box}(s,u) $ and $\textrm{Box}(u,t) $ and the three bubble integrals. The same conclusion also follows from solving integration-by-parts identities for scalar integrals using LiteRed \cite{Lee:2013mka, Lee:2012cn}.
The key observation to compute the full amplitude is that the only integrals appearing on the cuts are the three massless boxes and the three bubbles. Hence, one can construct an Ansatz,
\begin{equation}
A^1_4(s,t) = \lambda^4  \sum_{i \in \{s,t,u\} } (\alpha_i(s,t,D) \frac{1}{s_i^2} \textrm{Bub}_i +  \beta_i(s,t,D) \textrm{Box}_i )
\end{equation}
where the Ansatz coefficients are mass-dimension zero. In this particular case one can prove that there cannot be functions missed by all unitarity cuts. In short, the scattering amplitude must have a fixed, integer mass dimension. The coupling constant has a non-integer mass dimension in dimensional regularisation, as remarked above, and appears here with a power $4$ instead of $2$ as at tree level. This implies the coupling constant must always be multiplied by a non-integer power of $s$, $u$ or $t$ after integration, since there are no other scales in this problem. Hence, unitarity cuts pick up all integrals in the case where the external legs are massless. Note massive matter would introduce an immediate exception to this rule: massive tadpole integrals do not appear on unitarity cuts, see for instance \cite{Badger:2017gta} for analysis in this case.

In the case of massless matter, the coefficients in the Ansatz can be fixed by unitarity cuts. In particular the $s$, $u$ and $t$-channel bubbles are fixed by cuts in the respective channels, while the box coefficients appear in two cut channels. The latter leads to a consistency condition. In the case at hand 
\begin{equation}\label{eq:cutsphicubed_simped}
\textrm{disc}_{s} A^1(s,t) = \lambda^4 \left(  \frac{20-7D}{D-4}\,\frac{1}{s^2}   \textrm{Bub}_{\textrm{cut}}(s)  + 2  \, \textrm{Box}_{\textrm{cut}}(s,t) + 2 \, \textrm{Box}_{\textrm{cut}}(s,u) \right)
\end{equation}
is obtained. The $t$ and $u$ channel cuts follow in this case directly from Bose-symmetry. 

There is already some interesting physics in this example. For instance, the $\frac{1}{s^2}$ unphysical double pole multiplying the UV divergent bubble graph is indicative of the need for a mass-correction counterterm. Fixing the location of the pole of the four point amplitude in a given channel to a specific value determines the physical mass. In this on-shell approach, the counterterm comes from expanding a tree level amplitude with a variable, coupling constant dependent mass. Extending the full computation even in this scalar case to massive particles is very interesting, but will be discussed in future work \cite{upcomingMass}.

The technique just introduced is not limited to the specific example scalar field theory, but immediately covers all of effective field theory for scalar fields. In each case, one should fix three and four point scattering amplitudes at tree level. This yields the unitarity cuts of the one loop level amplitude. Using integration by parts identities then allows one to reduce each cut to the same cut master integrals. This is easily implemented for instance using the LiteRed\cite{Lee:2013mka, Lee:2012cn} package.

\subsubsection{Gluon loop contribution, adjoint external scalars}
Next consider minimally coupled adjoint massless scalar matter coupled to gluons, where the quadruple scalar self-interaction is not present. Here the colour factors will be consistently stripped off by employing colour-ordered amplitudes and focussing on leading colour. The extension to sub-leading colour is left as a straightforward exercise. The four point colour-ordered tree level scattering amplitude with two gluons and two scalars is
\begin{multline}\nonumber
A^0_{4}(1_g,2_g,3_s,4_s)  = {g_{\textrm{ym}}^2} \frac{2}{s t} \left[ -2 t (p_1 \cdot \xi_2) (p_3 \cdot \xi_1) +  2 (s+t) (p_2 \cdot \xi_1 )(p_3 \cdot \xi_2)  \right. \\ \left. +  2 s (p_3 \cdot \xi_1)( p_3 \cdot \xi_2 ) - (s+t) t (\xi_1 \cdot \xi_2) \right]
\end{multline}
Here $g_{\textrm{ym}}$ is the coupling constant of pure gluon interactions and mass-dimensionless. The four-scalar amplitude at one loop can be computed using the same strategy as above. The discontinuity in the s-channel for the leading trace colour-ordered amplitude involving gluon exchange reads
\begin{equation}
\textrm{disc}_{s, \textrm{gluon ex.}} A^1(s,t) = \int d\textrm{LIPS}_2 \sum_{\textrm{int. hel.}}A^0(1_s,2_s,L_{1,g}, L_{2,g}) A^0(3_s,4_s,-L_{1,g}, -L_{2,g})
\end{equation}
where the sum ranges over all internal helicities. Using the explicit expression for the amplitude as well as the completeness relation for the summed helicity indices yields
\begin{equation}\label{eq:cutglueloopfourscalar}
\textrm{disc}_{s\, \textrm{gluon ex.}} A^1(s,t) = g_{\textrm{ym}}^4\int d\textrm{LIPS}_2 \frac{1}{s^2 t_L t_R} \left[ f_g(t_L, t_R, s,t, D)\right]
\end{equation}
where $f_g$ is a function of mass-dimension $8$, polynomial in its variables. The explicit space-time dimension dependence in the function $f$ enters only through 
\begin{equation}
\sum_{\textrm{hel.}} \xi^{\mu}_{L_{1}} \xi^{\nu}_{L_{1}} \eta_{\mu \nu} = D-2.
\end{equation}
The result for the unitarity cut can in this case directly be re-interpreted as colour-ordered cut box, triangle and bubble integrals, where the cut is obtained by Cutkosky's rule\footnote{The reason for pointing this out explicitly is that it is general not true, see the two loop computation below.}. A difference to the scalar loop case in the $\phi^3$ case is the appearance of non-trivial loop momentum dependence in the numerator. Expanding the function $f$ in terms of the loop momentum dependency gives
\begin{eqnarray}
f_g(t_L, t_R, s,t)  &=& f_{0,0}(s,t) + \nonumber\\
&  & f_{1,0}(s,t) \, t_L +f_{2,0}(s,t) \, t_L^2 + f_{0,1}(s,t) \, t_R + f_{0,2}(s,t) \, t_R^2\nonumber\\
& & +  f_{1,1}(s,t)\, t_L t_R  +  f_{1,2}(s,t) \,t_L t_R^2 + f_{2,1}(s,t)\, t_L^2 t_R +\, f_{2,2}(s,t) t_L^2 t_R^2
\end{eqnarray}
The first line only contains a single cut box function, while the second line contains two cut triangle functions. The last line contains purely bubble-type contributions with numerators. Beyond the box functions, all other functions can be related by cut integration-by-parts identities to a cut $s$-channel bubble integral. This brings in another dependency on the space-time dimension, beyond that introduced by summing over internal polarisations in general. In this subsection they will be taken to be the same - this is part of a choice of regularisation scheme. Scheme dependence is explored more fully below.  

The identification of $f_{0,0}$  as the box coefficient also  follows from a $D$-dimensional quadruple cut \cite{Britto:2004nc} argument, which would set $t_L=t_R=0$. The loop momentum in this approach does not have to be explicitly parametrised as would be the case when applying spinor helicity methods. The computation for the $t$-channel cut of the same colour-ordered one loop amplitude is very similar. This gives the cut of the same box function as before ($\textrm{Box}(s,t)$)  as well as the cut $t$-channel bubble. 

Having taken care of the gluon exchange graph, what remains is the contribution of the scalar exchange contributions to the unitarity cut. The required cyclic four point tree amplitude reads
\begin{equation}
A^0_{4}(1_s,2_s,3_s,4_s)  = {g_{\textrm{ym}}^2} \left(\frac{u-t}{s}+\frac{u-s}{t} \right)
\end{equation}
This has to be fed into the scalar exchange contribution to the $s$-channel unitarity cut, leading to 
\begin{equation}\label{eq:cutscalarsloopfourscalar}
\textrm{disc}_{s, \textrm{scalar ex.}} A^1_{4}(1_s,2_s,3_s,4_s) =  4 {g_{\textrm{ym}}^4} \int d\textrm{LIPS}_2 \frac{1}{s^2 t_L t_R} \left[(s^2+s\, t_L+t_L^2) (s^2 +s\, t_R+t_R^2)\right]
\end{equation}
Note that the space-time dimension does not make an appearance in this computation: the particles on the cuts are after all scalars.

A full Ansatz for the one-loop colour-ordered four adjoint scalars amplitude can now be formulated. The Ansatz will consist of a single planar box as well as the $s$ and $t$ channel bubble integrals, with the coefficients functions of $s$, $t$ and $D$. Since these are the known master integrals of this topology, this Ansatz is certainly sufficient to capture the full amplitude. Using the cuts gives
\begin{eqnarray}
A^{1,\textrm{co}}(1,2,3,4) &=& {g_{\textrm{ym}}^4} \bigg[4 (s^2+t^2)\, \textrm{Box} (s,t)+\frac{(D^2-12D+16)\,s+18(D-4)\,t}{(D-4)\,s} \textrm{Bub}(s)\nonumber\\
&&\qquad\qquad+\frac{(D^2-12D+16)\,t+18(D-4)\,s}{(D-4)\,t} \textrm{Bub}(t)\bigg]
\end{eqnarray}
Note that without adding the scalar exchange contributions to the unitarity cut, the $t$ and $s$ channel cuts would be inconsistent with cyclicity. 

\subsubsection{Gravity loop contribution}

As a final one loop example with external scalars consider massless scalars, minimally coupled to gravity. The two-scalar, two-$\mathcal{N}=0$-multiplet amplitude reads:
\begin{equation}
M^0(1_G, 2_G, 3_s, 4_s) = \kappa^2 s\,A^0(1_g,2_g,3_s,4_s)\,A^0(1_g,2_g,4_s,3_s)
\end{equation}
The gravitational amplitude follows from this by projecting on the traceless symmetric part of the $\mathcal{N}=0$ multiplet legs. Also needed is the four scalar amplitude with $\mathcal{N}=0$-multiplet exchange,
\begin{equation}
M^0(1_s, 2_s, 3_s, 4_s) = \kappa^2 \frac{(s^2+t^2+u^2)^2}{s t u}
\end{equation}

The cut in the $s$-channel reads for $\mathcal{N}=0$-exchange 
\begin{equation}
\textrm{disc}_{s, \mathcal{N}=0\textrm{ ex.} } M^1(s,t) = \int d\textrm{LIPS}_2 \sum_{\textrm{int. hel.}} M^0(1_s,2_s,L_{1,G}, L_{2,G}) P_1 P_2 M^0(3_s,4_s,-L_{1,G}, -L_{2,G})
\end{equation}
where $P_1$ and $P_2$ indicate possible P-matrices selecting the internal quantum particles. The sums over internal helicities can be done with some effort in all cases. Here focus only on the simplest case with the whole  $\mathcal{N}=0$ multiplet in the loops. In this case the computation of the internal helicity sums factorises. Essentially this gives a double copy construction for the cut of the integrals. 
\begin{equation}
\textrm{disc}_{s} M^1(s,t) = \int d\textrm{LIPS}_2 \frac{1}{s^2 t_L t_R u_L u_R} f_{L'}(t_L, t_R, s,t, D) f_{R'}(t_L, t_R, s,t, D) 
\end{equation}
where the primed and unprimed $L$ and $R$ refer to factors in the double copy and sides of the unitarity cut respectively. The function $f$ is the proportional to the function encountered in \eqref{eq:cutglueloopfourscalar}. What is different is that the integrand does not look like that of a cut integral directly: there are too many propagators. However, expanding out the polynomials shows that only those terms that correspond to cut integrals remain. This is a consequence of locality of the tree level expressions. The scalar exchange contribution factorises trivially and can be added to the above contribution. 

IBP reduction of the results is straightforward and two cut box integrals ($\textrm{Box}(s,t)$, $\textrm{Box}(s,u)$), as well as a single cut bubble integral ($\textrm{Bub}(s)$) are obtained. Now write an Ansatz for the full amplitude as a sum over the three potential box integrals and the three bubble integrals,
\begin{equation}
M^1_4(s,t) = \kappa^4  \sum_i (\alpha_i(s,t,D) \textrm{Bub}_i +  \beta_i(s,t,D) \textrm{Box}_i )
\end{equation}
Taking the $s$-channel unitarity cut, properly IBP reduced then fixes the coefficients of the s-channel bubble, as well as those of the $(s,t)$ and $(s,u)$ channel box functions. The cuts in the $t$ and $u$ channel would fix the other variables, but these can be computed much more easily from demanding Bose-symmetry of the full result.  

The coefficients of the box type master integrals are
\bea
\beta_{s,t,D} = 16(s^4+t^4),\quad \beta_{s,u,D} = 16(s^4+u^4), \quad \beta_{t,u,D} =16(t^4+u^4),
\eea
where the lower indices stands for different ordering of external particles, i.e.,  $\beta_{s,t,D}$ is the coefficient of box master integral with ordering $(1,2,3,4)$, $\beta_{s,u,D}$ is for $(1,2,4,3)$ and $\beta_{t,u,D}$ is for $(1,3,2,4)$. The coefficients of the bubble type master integrals are much more lengthy.

\subsection{Two gluons, two scalars}
The next step toward more physical amplitudes is to allow for external spinning particles. First consider the case with two gluons and two adjoint scalars in the minimal coupling scenario. To compute the planar colour-ordered one loop amplitude, the tree level four point scattering amplitudes with four scalars, two scalars and two gluons and the one with four gluons are needed. There are two contributions to the unitarity cut:  either scalar or gluon exchange. The gluon exchange contribution reads
\begin{equation}
\textrm{disc}_{s, \textrm{gluon ex.}} A^1(1_s,2_s,3_g,4_g) = \int d\textrm{LIPS}_2 \sum_{\textrm{int. hel.}} A^0(1_s,2_s,L_{1,g}, L_{2,g})A^0(3_g,4_g,-L_{1,g}, -L_{2,g})
\end{equation}
The sum over internal helicities can be performed as before, but now the result has non-trivial external helicity dependence. By on-shell gauge invariance there must be coefficients such that 
\begin{equation}
A^1(1_s,2_s,3_g,4_g) = \sum_{i=1}^2 \beta_i B_i
\end{equation}
for the two linear independent solutions to the on-shell gauge invariance constraints. Two basis elements can be chosen as
\begin{equation}
B_1  =  - 2 \, (p_1 \cdot \xi_3)(p_3 \cdot \xi_4) - 2 \, (p_2 \cdot \xi_3)(p_3 \cdot \xi_4)- s (\xi_3 \cdot \xi_4)
\end{equation}
and
\begin{multline}
B_2  = 2 \,t\, (p_1 \cdot \xi_3) (p_2 \cdot \xi_4) + 2 \,(s+t) \,(p_2 \cdot \xi_3) (p_2 \cdot \xi_4) + 2\, (s+t) \,(p_2 \cdot \xi_3)(p_3 \cdot \xi_4)  \\  - 
t ( s +t) \, (\xi_3 \cdot \xi_4)
\end{multline}
according to eq.(\ref{basis:2gluon2scalar-1}) and eq.(\ref{basis:2gluon2scalar-2}).

To obtain the scalar functions $\beta_i$, the P-matrix method is used. First compute
\begin{eqnarray}\label{eq:afterextsumI}
&&\sum_{\textrm{ext. hel.}}  B_i\,\, \textrm{disc}_{s, \textrm{gluon ex.}}  A^1(1_s,2_s,3_g,4_g) \nonumber\\
&&\qquad= \int d\textrm{LIPS}_2\sum_{\textrm{ext. hel.}} B_i \left(\sum_{\textrm{int. hel.}} A^0(1_s,2_s,L_{1,g}, L_{2,g})A^0(3_g,4_g,-L_{1,g}, -L_{2,g})\right)  
\end{eqnarray}
The sums over internal and external helicities can be performed consecutively. A subtle issue here concerns a choice of renormalisation scheme and a choice of calculational scheme. The dimension of space-time enters in three separate places. For the external and internal helicity sums one can take
\begin{align}
\sum_{\textrm{ext. hel.}} \xi^{\mu}_{p_{i}} \xi^{\nu}_{p_{i}} \eta_{\mu, \nu} &= D_{\textrm{ext}} - 2 \\
\sum_{\textrm{int. hel.}} \xi^{\mu}_{L_{1}} \xi^{\nu}_{L_{1}} \eta_{\mu, \nu} &= D_{\textrm{int}} - 2
\end{align}
and the dimension enters into the solution of the IBP relations as the dimension of the loop integrals,  $D_{\textrm{loop int.}}$. The choice of external dimension is part of a choice of renormalisation scheme: natural values are four or $D_{\textrm{loop int.}}$. The internal dimension is a choice of calculational scheme: if one wants to use four dimensional unitarity cuts (which miss some information on the full amplitude), one should set this to four. In this article we will use the more physically natural value of $D_{\textrm{loop int.}}$ for the internal dimension. Moreover, we will first choose the external dimension to be the same so that 
\begin{equation}
D_{\textrm{loop int.}} = D_{\textrm{ext}} = D_{\textrm{int}} = D\qquad 
\end{equation}
holds unless explicitly stated otherwise. Choosing different values for the dimensions subtly breaks manifest $SO(D)$ invariance into different Lorentz groups. This is a complication we first want to avoid, postponing a full discussion to the four gluon case below. With renormalisation conventions fixed equation \eqref{eq:afterextsumI} shows that in the case at hand two computations have to be performed, one for each choice of basis element $B_i$. For each of these two the computation is structurally exactly the same as before since after summation over all helicities only a scalar function remains. 
In this special case the master integrals involve only box and bubble types and IBP on the cuts can be used to isolate the coefficients of box and s-channel bubble.

By inverting the P-matrix of the gauge-invariant basis elements, coefficients of $B_i$ from this cut can be derived.  Specifically, with the Ansatz 
\begin{eqnarray}
\textrm{disc}_{s, \textrm{gluon ex.}} A^1(1_s,2_s,3_g,4_g) =  \beta_{1, s, \textrm{gluon ex.}}B_1 +\beta_{2, s, \textrm{gluon ex.}} B_2
\end{eqnarray}
projected onto basis elements $\{B_1,\, B_2\}$ by summing over external helicities, inverting the P-matrix from IBP on cuts results, the coefficients of basis elements read 
\begin{eqnarray}
\beta_{i, s, \textrm{gluon ex.}}= (P^{-1})_{ij}(c_{\textrm{Box},j}\textrm{Box}(s,t)+c_{\textrm{Bub(s)},j}\textrm{Bub}(s)).
\end{eqnarray}

Similarly, the scalar exchange contribution reads
\begin{equation}
\textrm{disc}_{s, \textrm{scalar ex.}} A^1(1_s,2_s,3_g,4_g) = \int d\textrm{LIPS}_2 A^0(1_s,2_s,L_{1,s}, L_{2,s}) A^0(3_g,4_g,-L_{1,s}, -L_{2,s}).
\end{equation}
The coefficients $\beta_{i, s, \textrm{scalar ex.}}$ for gauge invariant basis elements $B_1$ and $B_2$ from this contribution can be derived as above. As before the Ansatz for this one-loop colour-ordered two adjoint scalars and two gluons amplitude will consist of a single planar box as well as the $s$ and $t$ channel bubble integrals, with the coefficients functions of $s$, $t$ and $D$. Proceeding as before gives the final result as
with
\begin{eqnarray}
\beta_1 &=& -\frac{2 (D-4) s t}{(D-3) u} \textrm{Box}(s,t) -\frac{8}{u} \textrm{Bub}(t)\nonumber\\
&&+\frac{\left(D^2-12 D+16\right) s^2+\left(D^2-2 D-8\right) s t+2 (D-4) t^2}{(D-2) s^2 u }\textrm{Bub}(s)
\end{eqnarray}
and
\begin{eqnarray}
\beta_2 &=& -\frac{2 (D-2) s^2+4 (D-3) s t+4 (D-3) t^2}{(D-3) u^2} \textrm{Box}(s,t,D)\nonumber\\
&&-\frac{8 (D-6) s+16 (D-4) t}{(D-4) s u^2} \textrm{Bub}(s) +\frac{8 (D-2)}{(D-4) u^2} \textrm{Bub}(t) .
\end{eqnarray} 
In the computation, s- and t-channel cuts share the same coefficients of $\textrm{Box}(s,t,D)$ for each $B_i$ while the coefficients of $\textrm{Bub}(s)$ and  $\textrm{Bub}(t,D)$ should be determined separately. The unphysical apparent pole in $u$ appearing in the final results is a spurious singularity for this colour-ordered amplitude. Its residue can be checked to vanish from the explicit formula for the massless box function in \cite{Smirnov:2004ym} as well as for the bubble integrals. Collecting terms proportional to $s^{\epsilon}$ and $t^{\epsilon}$ shows the cancellations when expanding in $u$.

\subsection{Four gluons}

Next up is the four gluon amplitude at one loop. Before continuing, let us first consider colour quantum numbers. Re-instating colour, the s-channel cut reads
\begin{equation}
\textrm{disc} A^1_4(1,2,3,4) = \int d\textrm{LIPS}_2  \sum_{\textrm{colour, internal helicities}} A^0_4(1,2,l_1,l_2) A^0_4(l_1,l_2,3,4) 
\end{equation}
as illustrated in the figure. The Lorentz-invariant phase space integral and sums over helicities involve the unobserved particles in the loop process. It is advantageous to use the D${}^3$M \cite{DelDuca:1999rs} basis to express the colour dependence of the tree amplitudes on the cut. This basis requires singling out two legs. On a double cut at one loop it is natural to pick the loop integral legs to be these legs. The full colour tree amplitude can be expressed in the  D${}^3$M basis as sum over permutations of the colour-ordered amplitudes multiplied with a string of contracted structure constants,
\begin{equation}
A^{\textrm{full}}(l_1, 2,3, \ldots,j,  l_2) = \sum_{\sigma \in P(2,3,\ldots, j)} A^{\textrm{co}}(l_1, \sigma, l_2) f^{a_{l_1} a_{\sigma_1} a_{i_1}} f^{a_{i_1} a_{\sigma_2}, a_{i_2}} \ldots f^{a_{i_{j-1}} a_{\sigma_{j}}, a_{l_2}}
\end{equation}
When inserted on a double cut, it is obvious that the string of structure constants becomes the circle of structure constants which is the one-loop version of the  D${}^3$M basis. Hence, in this basis colour is straightforward to include, and simply reduces to studying colour-ordered amplitudes, as is well known. 

Given colour-ordered tree amplitudes, one can construct the discontinuity of the next order in perturbation theory, the one loop correction by the formula above. Since the integrated one loop four gluon amplitude for instance obeys the on-shell constraints, it can be expanded in the basis. The lower-order amplitudes which appear in the unitarity formulas are however still to be integrated: they therefore depend on integration momenta. The P-matrix method is ideally suited to obtain the integrand in a useful form. The reason is that after multiplication with a basis element and summation over the internal and external helicity quantum numbers the end result is a phase space integral over a function of only external and internal momenta. 
\begin{multline}
\textrm{disc} \sum_{\textrm{external helicities}} B_i(s,t) A^1_4(1,2,3,4) =\\ \int d\textrm{LIPS}_2  \sum_{\textrm{all helicities}} B_i(s,t) A^0_4(1,2,l_1,l_2) A^0_4(l_1,l_2,3,4.) 
\end{multline}
The right hand side is after summation over helicities by using the completeness relations a scalar-type integral, instead of the tensor-type integral before. This is the same mechanism as first advocated in \cite{Glover:2003cm} to simplify sums over Feynman graphs, applied here to unitarity contributions for the first time. The on-shell conditions for the internal momenta enforced by taking a unitarity cut are easily implemented. They require only consideration of inner products of the internal momentum with other momenta - no explicit solution for the internal on-shell momentum is needed and the number of dimensions need not be specified. Put differently, we employ Baikov variables. 

For the case at hand, there are after momentum conservation on both the left and right factor of the cut only four independent inner products of the four independent momenta involved after momentum conservation, say the set $p_1,p_2, p_3$ and $l_1$. Beyond the external Mandelstam variables $s, t$ these can be taken to be  $t_L$ and $t_R$: the Mandelstam variables of the left and right amplitudes on the unitarity cut. Solving the on-shell conditions and momentum conservation then yields 
\begin{align}
l_1 \cdot p_1 & = -\frac{s + t_L}{2}\\
l_1 \cdot p_2 & =  \frac{t_L}{2}\\
l_1 \cdot p_3 & = - \frac{t_R }{2}\\
l_1 \cdot l_1 & = 0
\end{align}
For a quadruple cut, one should simply take a double residue for $t_L$ and $t_R$ to zero. Since the pole is manifest this is a simple computation. The result of the summations over internal and external helicities is an explicit expression on the right hand side 
\begin{equation}
\textrm{disc} \sum_{\textrm{external helicities}} B_i(s,t) A^1_4(1,2,3,4) = \int d\textrm{LIPS}_2\,\,  I_i (s,t,t_L, t_R)
\end{equation}
The integrand on the right hand side only contains single particle singularities in the invariants $s, t_L$ and $s, t_R$ respectively by lower order-unitarity. Structurally the internal momentum dependence of the obtained expression reads
\begin{equation}
I_i (s,t,t_L, t_R) \propto \sum_{j=-1, k=-1}^{3,3} c_{ijk}(s,t)  t^j_{L} t^k_{R}
\end{equation}
The general class of cut loop integrals can be reduced by IBP reduction to an IBP basis involving only the cut loop integrals. See \cite{Larsen:2015ped} \cite{Ita:2015tya} and \cite{Georgoudis:2016wff} for promising developments into solving IBP relations on cuts. Here we can take a simpler approach: temporarily re-instating the cut propagators yields a class of box-type integrals with numerators, 
\begin{equation}
 \sum_{j=0, k=0} c_{ijk} \int d^Dl \frac{1}{(l_1)^2 (l_1+p_1+p_2)^2(l_1+p_2)^2(l_1-p_3)^2} \left(  ((l_1-p_3)^2)^j ((l_1+p_2)^2)^k \right)
\end{equation} 
The numerators simply start cancelling propagators in the loop integral. This yields in this simple case triangle integrals with two on-shell legs, which can be integrated easily using only textbook methods. Additional numerators in the triangles are similarly easy to treat. To automatise this part of the computation it is more convenient (and completely equivalent) to treat this as solving the IBP problem without cuts using public codes such as FIRE \cite{Smirnov:2014hma}, LiteRed \cite{Lee:2013mka} or Reduze \cite{vonManteuffel:2012np}.  This reduces every triangle-type integral as a bubble integral with the singularity that is picked out by the unitarity cut. In more general situations than one loop amplitudes with four massless external particles, one aims to solve the IBP problem in the classes of integrals to arise in all unitarity cuts. This expresses all integrals in a basis of integrals known as master integrals. Then one can cut the resulting integral reduction expressions to obtain IBP reductions of the phase space integrals. The upshot is an expression 
\begin{equation}
\textrm{disc} \sum_{\textrm{external helicities}} B_i(s,t) A^1_4(1,2,3,4) = \sum_{j=1}^{\textrm{nr. cut masters}} d_{ij} \textrm{MI}_{j}
\end{equation}
In the one loop planar four point amplitude case there are three boxes and three bubble master integrals. There are only two cut master integrals appearing on the right hand side in a given channel: a cut scalar box integral and a cut scalar bubble integral. Hence the total one-loop amplitude is expressible in terms of a scalar box and two scalar bubbles for planar colour-ordered gluon amplitudes. The coefficients of these expressions are functions only of the external kinematics, as well as the dimension. There is a double basis expansion: the basis for external kinematics captured by solving the on-shell constraints and the basis for the master integrals,
\bea
A^1_4(1,2,3,4) = \sum_{i,j} \, c_{ij}(s,t,D)\, \textrm{MI}_j B_i.
\eea
The coefficients can be computed by unitarity and are read off rather straightforwardly from the expressions after implementing IBP relations on the cuts. Coefficients multiplying functions that have cuts in more than one channel, such as the scalar box, must match between channels. The box coefficient in the explicit example here can be most easily computed by taking a quadruple cut, setting $t_L$ and $t_R$ to zero in the above computation. This observation guarantees consistency between $s$ and $t$ channel cuts. 
 
A subtlety concerns the choice of regularisation scheme. So far all momenta and polarisations both internal and external were treated in $D$ dimensions. Other choices can be made easily by using the appropriate completeness relations. Some care has to be taken for the metric appearing in these relations: these are either $D_{int}$ or $D_{ext}$ dimensional. This plays a role whenever the internal momentum is involved, where one should decompose
\begin{equation}
l^{D}_{\mu} = l^{4}_{\mu} + l^{-2 \epsilon}_{\mu} 
\end{equation}
The internal states have to be $D$-dimensional. Poincar\'e invariance implies that after summing over helicities the D-dimensional part of the momentum can only occur as $\mu^2 = (l^{-2 \epsilon})^2 $ in the integrand. Computations have been done for the scalar-in-loop and pure gluon cases. The coefficient tables $c_{ij}$ of ten gluon basis elements and master integrals can be found in the auxiliary file.

One feature of the result is the presence of unphysical poles in coefficients. For a given choice of basis, the coefficients $\beta_i$
\begin{equation}
A^1 = \sum_i \beta_i B_i
\end{equation}
can separately have for instance tree level poles for certain momenta. As already anticipated above, there can be relations between basis elements $B_i$ on singular loci, i.e. on the roots of some polynomial equation $f(s,t)=0$,
\begin{equation}
\left(\sum_{i} \alpha_i B_i \right)_{f(s,t)=0}
\end{equation}
Typically, this type of singularity in coefficients involves the roots of the completely anti-symmetric polynomial. A simple way to check if poles in coefficients are cancelled in the full result is to insert the vector form of the tensors $B_i$ in the space spanned by solutions to the Poincar\'e constraints and transversality (the space of the `$T$' matrices above). 

A second type of singularity does not appear because of a poor choice of basis for the tensor structures, but because of the choice of integral basis. For instance, the box integral coefficient displays up to $\frac{1}{u^4}$-type singularities. This type of spurious singularities in box integral coefficients is known, see e.g. \cite{BjerrumBohr:2007vu}, \cite{Bern:1995db}, but not oft discussed. For the colour-ordered gluon scattering amplitude to make sense it cannot contain any singularity in the u-channel. Hence, the first four coefficients of the Taylor expansion of the box integral of $u$ around $u=0$ must cancel against bubble integrals in both the $s$ and the $t$ channel. We have checked explicitly that this indeed happens, in $D$ dimensions. In the process we note that the bubble integral coefficients are determined up to a single function of $D$ by these constraints. This is because the $s$ and $t$ channel bubble integrals can be multiplied by $\frac{1}{s^2}$ and $\frac{1}{t^2}$ respectively for generic IR singularities at this loop order, and the coupling constant is dimensionless in four dimensions.

\subsubsection{Basis choice for IBP reduction}
In the computation above the IBP relations have been used to simplify the product of tree amplitudes. There are two choices involved in this computation: a choice of basis for solving the IBP relations as well as a choice of which IBP relations to solve. The first is essentially a basis change which will not be exploited here. The second will be used to reveal some interesting physics. In particular, in the usual approach to one loop amplitudes one uses a basis for the loop amplitudes which consists of box, triangle and bubble integrals, as well as these types of integrals with so-called $\mu^2$ type terms, where $\mu^2$ is the square of the part of the loop momentum pointing in the $-2\epsilon$ dimensions of dimensional regularisation. In the usual approach to loop amplitudes the coefficients of this basis are $\epsilon$-independent, four dimensional expressions.

It is an interesting problem to obtain an IBP reduction for all integrals which appear at the one loop cuts to the basis with $\mu^2$. Although this could be studied directly in principle, in practice it is more convenient to obtain this reduction using the automatic tools at our disposal. The key clue is the D-independence of the coefficients in the basis expansion used often in the literature. Hence, we are looking for an IBP reduction where the reduction coefficients for the scalar integrals appearing on the cut are dimension-independent. An algorithm to obtain a dimension-independent IBP reduction having calculated a dimension-dependent one before is described in \cite{Boels:2016bdu}. One first obtains an expression for all the integrals of interest in a particular problem  in terms of the master integrals,
\begin{equation}
I_j = \sum_{i} c_{j,i}(s,t,\epsilon) \textrm{MI}^i
\end{equation}
using the full IBP reduction. Interpret the index on the master integrals $MI$ as a row index and the index on the original integrals $I$ as a column index. Then evaluate this reduction for $c$ different, random integer values for the dimension, creating a matrix has the number of $I$ integrals as columns and $c$ times the number of master integrals as the number of rows. The kernel of this matrix spans the set of dimension-independent IBP relations between the integrals $I$ in the set, as long as $c$ is large enough. In general, the number of rows should exceed the number of columns, setting a lower bound on $c$. In this approach, there is a chance accidental relations arise due to matrix degeneracies for poorly chosen random integer values for the dimension. These typically lead to relations with very large rational numbers: this forms a practical low-level consistency check on the outcome. 

For the case at hand one first needs to express all the integrals in the same set of dimension dependent master integrals. A roadblock is that $\mu^2$-containing integrals are naturally related to scalar integrals in higher dimensions. In particular,
\begin{align}
\int d^{D}l \mu^2 \left( \ldots \right) &= - \epsilon \int d^{D+2}|  \left( \ldots \right)  \\
\int d^{D}| \mu^4 \left( \ldots \right) &= - \epsilon (\epsilon-1) \int d^{D+2}|  \left( \ldots \right) 
\end{align}
hold. LiteRed and most other public IBP solvers typically involve same dimension master integrals only. In other words, all integrals are assumed to be in the same number of dimensions. There are known relations between integrals in different numbers of dimensions, either increasing or decreasing the dimensions, see e.g. \cite{Tarasov:1996br}. These relations are coded into LiteRed and can be used to express the $\mu^2$ containing integrals in terms of the dimension-full IBP basis in $D$ dimensions with some work.

\subsubsection{Results in different schemes}
The scheme dependence of the results can be studied explicitly. The scalar loop results derived from unitarity cuts and P-matrix method can be further expressed in terms of spinor helicity formalisms and master integrals of a box and two bubbles, by requiring external scattering particles in four dimensions
{\allowdisplaybreaks
\begin{eqnarray}
&& A^{1,{\rm scalar}}(1^+,2^+,3^+,4^+)=\frac{2i}{ (4\pi)^{2-\epsilon}} \frac{- s \,t}{\langle 1,2\rangle \langle 2,3\rangle \langle 3,4\rangle \langle 4,1\rangle}\bigg[-\frac{(D-4) (D-2) s^2 t^2}{16 (D-3) (D-1) u^2} \textrm{Box}(s,t)\nonumber\\
&&\qquad\qquad\qquad\qquad-\frac{t (2 (D-3) s+(D-4) t)}{4 (D-1) u^2}\textrm{Bub}(s)-\frac{s ((D-4) s+2 (D-3) t)}{4 (D-1) u^2}\textrm{Bub}(t)\bigg]\nonumber\\
~~\\
%%%%%%%%%%%%%%%%%%%
&& A^{1,{\rm scalar}}(1^-,2^+,3^+,4^+)=\frac{2i}{ (4\pi)^{2-\epsilon}} \frac{[2,4]^2}{ [1,2]\langle 2,3\rangle \, \langle 3,4\rangle\,[4,1]} \frac{s\,t}{ u}\bigg[\frac{ D (D-4) s^2 t^2}{16 (D-3) (D-1) u^2} \textrm{Box}(s,t) \nonumber\\
&&\qquad\qquad -\frac{s \left(\left(D^2-16\right) s t+2 (D-4)
   s^2-8 t^2\right)}{4 (D-2) (D-1) t u^2}\textrm{Bub}(s)-\frac{t \left(\left(D^2-16\right) s t+2 (D-4)
   t^2-8 s^2\right)}{4 (D-2) (D-1) s u^2}\textrm{Bub}(t)\bigg]\nonumber\\
~~\\
%%%%%%%%%%%%%%%%%%%
&& A^{1,{\rm scalar}}(1^-,2^-,3^+,4^+)=2\frac{\mathcal A_4^{\rm{tree}}(1^-,2^-,3^+,4^+)}{ (4\pi)^{2-\epsilon}}\bigg[\frac{(D-4) (D-2) s t^3}{16 (D-3) (D-1) u^2} \textrm{Box}(s,t) \nonumber\\
&&\qquad\qquad\qquad\qquad\qquad\qquad\qquad -\frac{s (D t+2 s)}{4 (D-1) u^2}\textrm{Bub}(s)+\frac{t ((D-4) s+2 (D-3) t)}{4 (D-1) u^2} \textrm{Bub}(t)\bigg]\nonumber\\ 
~~\\
%%%%%%%%%%%%%%%%%%%
&& A^{1,{\rm scalar}}(1^-,2^+,3^-,4^+)=-2\frac{\mathcal A_4^{\rm{tree}}(1^-,2^+,3^-,4^+)}{ (4\pi)^{2-\epsilon}}
  \bigg[-\frac{D (D+2) s^3 t^3}{16 (D-3) (D-1) u^4} \textrm{Box}(s,t)\nonumber\\
&&\qquad+\frac{s \left(2 (D-4) (D-2) s^3+(D-4) (D-2) (D+8) s^2 t+16 (5-2 D) s t^2-8 (D-4) t^3\right)}{4 (D-4) (D-2) (D-1) u^4} \textrm{Bub}(s)\nonumber\\
&&\qquad+\frac{t \left(-8 (D-4) s^3+16 (5-2 D) s^2 t+(D-4) (D-2) (D+8) s t^2+2 (D-4) (D-2) t^3\right)}{4 (D-4) (D-2) (D-1) u^4} \textrm{Bub}(t)\bigg].\nonumber\\
\end{eqnarray}
}
These results are consistent with the results in the literature in \cite{Bern:1995db}. 

Rather than the scalar loop shown above one can also study loop contributions from the pure Yang-Mills theory, where the internal propagating particles are gluons. Results for both schemes are listed in the auxiliary file. Comparing these results shows that if one takes $D=4-2\epsilon$ as usual, results from two schemes share the same all-plus and single-minus helicity one-loop expressions while difference for MHV configurations are proportional to tree-level amplitudes, with a universal factor. This difference reconfirms that the two computations are really different renormalisation schemes. The results from FDH scheme agree with expressions by combining scalar-loop, $\mathcal{N}=1$ one-loop and $\mathcal{N}=4$ one-loop contributions \cite{Bern:1995db}. This is an important sanity check of the method. 

A feature of the computation which is interesting is the fact that the non-symmetric basis elements all come with vanishing coefficient. As analysed in appendix B, these coefficients only influence the one-helicity-unequal amplitudes. This motivates a general conjecture for which more evidence will be presented below: the non-symmetric basis element coefficients for the four gluon amplitudes always vanish, independent of scheme. The fact that these coefficients are zero at tree, one (and two) loop order for planar amplitudes restricts their form at higher loops and for non-planar amplitudes. Although the conjecture applies to all regularisation schemes, it is instructive to work out its consequences in the four dimensional helicity scheme. As shown in the appendix, the non-symmetric basis elements only contribute to one helicity un-equal amplitudes. Since these three are zero, the four in principle independent parity-even one helicity unequal amplitudes can be computed from the computation of one. Taking the expressions from the appendix, one arrives at explicit `helicity switching' identities:
\begin{equation}\label{eq:helswitchingGluons}
\frac{A(1^-, 2^+, 3^+, 4^+)}{\left([2,3] [2,4] \right)^{2}}  = \frac{A(1^+, 2^-, 3^+, 4^+)}{ \left([1,3] [1,4] \right)^{2}}
\end{equation}
Verification of the helicity switching conjecture to higher orders in either coupling constant or $\frac{1}{N_c}$ exceeds the scope of the present paper, as is the extension to multiple particles. Note the only amplitude relations at one loop for integrated expressions we are aware of in \cite{BjerrumBohr:2011xe} do not involve helicity switching. Instead, the helicity switching identities have much more in common with supersymmetric linear Ward identities, see e.g. eq.(77) in \cite{Dixon:1996wi}. It could be however that the derived relations are a consequence of a kinematic accident for four massless particles more than of a physical symmetry. In either case, there is a strong hint that Bose symmetry plays a big role in the explanation.

One interesting application of the tensor structure formalism is to study divergences of the amplitude.  Consider for instance the $\frac{1}{\epsilon}$ divergence in the epsilon expansion for $D=6-2 \epsilon$. Since there are no IR divergences in this number of dimensions, this is a straight-up UV divergence. Since this divergence can be checked to be non-zero and explicitly involves other tensor structure basis elements it is seen the theory is not renormalisable, at least not in the traditional sense (see however also \cite{Bork:2015zaa}). Similarly one can check that for $D=4-2\epsilon$ the maximal $\frac{1}{\epsilon^2}$ pole is proportional to the tree amplitude as required by generic considerations of IR divergences.

\subsection{Four gravitons}
In this subsection we consider the one-loop scattering amplitude for four gravitons and/or dilatons, with a scalar flowing in the loop. Like the gluon case one can work in either the D-dimensional or the 't Hooft-Veltman scheme of dimensional regularisation. Although no new computational ideas are needed, gravitons introduce a new level of algebraic complexity. Tthe kinematic basis of solutions to the on-shell constraints is $56$ dimensional for symmetric external matter in D-dimensions, and each of them is of a lengthy expression. In four dimensions there are only $8$ independent parity even graviton scattering amplitudes \eqref{sum-puregrav-4dim}. All eight independent strictly four-dimensional four-graviton amplitudes come from squared-gluon basis elements and thus computations simplify significantly. In the four dimensional external matter setup, there are results in the literature available \cite{Dunbar:1994bn}. Hence the focus here will be primarily on the four dimensional spinor helicity scheme.

\subsubsection{Scalar loop}
For minimally coupled massless scalars the needed two-scalar two-graviton amplitude reads 
\bea
M^0(1_G, 2_G, 3_s, 4_s) = \kappa^2 s\,A^0(1_g,2_g,3_s,4_s)\,A^0(1_g,2_g,4_s,3_s).
\eea
which can be obtained straightforwardly using a generalisation of the method used above to derive four gluon amplitudes. The cut in the s-channel of the one loop amplitude gives
\bea
\textrm{disc}_{s} M^1 (1_G,2_G,3_G,4_G)= \int d\textrm{LIPS}_2  M^0(1_G,2_G,L_{1,s},L_{2,s})\,M^0(3_G,4_G,-L_{1,s},-L_{2,s}).
\eea
An amplitude of four graviton can also be written in terms of gauge invariant basis elements and their corresponding coefficients
\bea
M^1 (1_G,2_G,3_G,4_G)=\sum_{i=1}^N \beta_i B_i.
\eea
using the graviton basis constructed above. 

In 4-dimensional space-time only eight of the ten gluon basis elements are independent and two sets of gluon basis elements in 4-dim can form $8\times 9/2=36$ left-right factorised graviton basis elements. The four non-double copied basis elements remain the same.

Rather than the $\mathcal N =0$ P-matrix, the pure graviton projection sums over external helicities as 
\bea\label{sum-puregrav-4dim}
       \sum \xi^{\mu\nu} (\xi^{\rho\sigma})^\dagger = (\mathcal P^{\mu\rho}\mathcal P^{\nu\sigma}+\mathcal P^{\mu\sigma}\mathcal P^{\nu\rho}-\frac{2}{ D-2}\mathcal P^{\mu\nu}\mathcal P^{\rho\sigma})
\eea 
where 
\bea
      \mathcal P^{\mu\nu} = \eta^{\mu\nu}- \frac{p^\mu q^\nu + p^\nu q^\mu}{p\cdot q}.
\eea
is the spin one polarisation sum. Projecting the cut-containing expressions onto graviton basis elements by summing over helicities of external particles via eq.(\ref{sum-puregrav-4dim}) with $D=4$ reads 
\bea
&&\sum_{\textrm{ext. hel.}} B_i\,\, \textrm{disc}_{s}  M^1 (1_G,2_G,3_G,4_G)\nonumber\\
&&\qquad =\int d\textrm{LIPS}_2 \sum_{\textrm{ext. hel.}} B_i \bigg( M^0(1_G,2_G,l_{1,s},l_{2,s})\,M^0(3_G,4_G,-l_{1,s},-l_{2,s})\bigg).
\eea
In this projection the Lorentz products of the D-dim loop momentum with external momenta in 4-dim are treated as was done above for the pure gluon.  With 4-dimensional Minkowski metric, scalar products of cut loop momentum satisfy $l_{i,S}\cdot L_{i,S}= \mu^2$. The s-channel cut onto 4-dimensional graviton basis elements contains integrands up to $\mu^8$. Since $\mu^2$ dependent integrals are equivalent to integration in higher dimension, dimension recurrence relations and IBP relations can reduce all cut integrals into box and bubble type master integrals. 

From s-channel cut onto graviton basis elements, coefficients of box-type master integrals $\textrm{Box}(s,t)$ (ordering of external particles $(1,2,3,4)$) and $\textrm{Box}(s,u)$ (ordering of external particles $(1,2,4,3)$) as well as the coefficient of bubble-type master integral $\textrm{Bub}(s)$ are derived.  Together with cuts from t-channel and u-channel, the four-graviton scalar loop example results in
\begin{equation}
M^1 (1_G,2_G,3_G,4_G) = \kappa^4  \sum_i (\alpha_i(s,t, D) \textrm{Bub}_i +  \beta_i(s,t,D) \textrm{Box}_i ).
\end{equation}
Coefficient tables of eight independent graviton basis elements and three master integrals can be found in the auxiliary file. Note that from the s-channel cut one can derive the coefficients of the other cuts by Bose symmetry in this case. Translating the basis elements into spinor-helicity formalism with specific helicity configurations, all-plus, single-minus and MHV amplitudes agree with \cite{Dunbar:1994bn} up to an overall convention-dependent numerical factor. Detailed expressions are also given in the auxiliary file.

Strikingly, from inspecting the answer it follows readily that out of eight basis elements, three come with zero coefficient for any of the master integrals. From counting alone this is reminiscent of the gluon helicity shifting identity discussed above. It can indeed be checked that 
\begin{equation}\label{eq:helswitchingGravitons}
\frac{M(1^-, 2^+, 3^+, 4^+)}{\left([2,3] [2,4] \right)^{4}}  = \frac{M(1^+, 2^-, 3^+, 4^+)}{ \left([1,3] [1,4] \right)^{4}}
\end{equation}
holds in the four dimensional helicity scheme. Again we conjecture this is an all-loop relation; it would be interesting to verify it at two loops.

\subsubsection{On more general matter in the loop}

The scalar-in-the-loop computation can also be performed within the D-dimensional renormalisation scheme. The algebraic complexity of this problem increases due to the size of the tensor structure basis both for the integrand as well as for the external amplitudes. Ultimately, this is simply due to the number of different polarisation vectors, which leads to a rather quick growth of the number of needed calculational steps within the used formalism. Nevertheless, we have been able to push this computation through for $\mathcal{N}=0$ as well as symmetric internal matter.

The biggest bottleneck is the inner product with the external basis structures. Memory use in intermediate steps remains a major issue, but in our current setup the computation fits into that of a laptop (a few GB). The result shows that all basis elements which involve a (left-right symmetrised) double copy with at least one of the non-symmetric gluon elements vanishes, but only after IBP reduction. In addition, the coefficients of three of the four non-squaring elements vanish, leaving only the Bose-symmetric element in this sector. This coefficient is not zero however. Note that this pattern is highly reminiscent of the result in the gluon sector: the tensor elements which cannot be Bose-symmetrised are zero. We conjecture this pattern will hold generically for general matter at higher loop orders. In the gluon case, it was shown above that the vanishing of basis element coefficients leads to relations between amplitudes in the four dimensional helicity scheme. In fact, since the coefficients of the graviton basis vanish for a set of double copy derived elements and since graviton amplitudes of $4$-dimension do not involve the non-double copy basis elements at all, it is not so difficult to motivate equation \eqref{eq:helswitchingGravitons}. 

The IR structure of amplitudes with gravitons and dilatons is a prime candidate for further research, see e.g. \cite{Weinberg:1965nx} for inspiration. This we leave for future work, besides noting the general absence of $\frac{1}{\epsilon^2}$ poles in our expressions as expected. One obvious point for improvement is in our choice of basis; from the explicit expressions (see the attached notebook for the scalar in the loop case) it is clear that the coefficients of the basis elements have many spurious singularities, e.g. poles in $s-t$ as well as higher order poles. Some cancel in the sum over the different basis elements. Especially singularities which are contained in completely symmetric polynomials are expected to be reducible by a better choice of basis. Apart from aesthetics, this should lead to smaller intermediate and final expressions for the basis coefficients.

\section{Four gluons at two loops, planar case}
\label{sec:twoloop4pt}

As a next step toward collider-relevant observables, let us consider the two loop planar four point gluon scattering amplitude in pure Yang-Mills theory. This amplitude was computed in \cite{Bern:2002tk} in the four dimensional helicity scheme, see \cite{Abreu:2017xsl} for a recent re-computation. 

Results will be presented here first in the regularisation scheme where internal and external gluons are in D dimensions, as this is most natural for tensor structures.  The maximal set of IBP relations will be used, leading to a minimal IBP basis. The internal and external tensor dimensions are more-or-less trivial to generalise following the one-loop analysis above. Results in 't Hooft-Veltman regularisation scheme will also be given, where all identical helicity ones have been cross-checked to those in \cite{Bern:2000dn}.

\subsection{Integral basis and cuts of integrals}
The needed integral topologies can be detected by considering the singularities of the triple cut in the $t$ channel. On this cut,
\begin{multline}
\textrm{cut} A^{2}(1,2,3,4) = \int DLIPS_3 \sum_{\textrm{int. hel.}}A_5^0(2,3,l_1,-l_1 - l_2 - p_2 - p_3, l_2) \\ A_5^0(4,1,-l_1, l_1+l_2 + p_2+p_3, -l_2) 
\end{multline}
which includes two five gluon colour ordered Yang-Mills amplitudes. First list all the possible poles of this expression
\begin{equation}
\{s,t, (l_1 + p_3)^2,  (l_2 + p_2)^2, (l_1 - p_4)^2, (l_2 -p_1)^2, ( l_2 + p_2 + p_3)^2, (l_1 + p_2 + p_3)^2 \}
\end{equation}
where the last two expressions can occur quadratically. Note there are six loop momentum-containing propagators. Reinstating the cut propagators this gives an ordered list of $9$ propagators which forms a complete basis for all Lorentz products of two momenta from the set $\{l_1, l_2, p_1, p_2, p_3\}$. The momenta in these propagators can be chosen to be:
\begin{equation}\label{eq:integralbasisplanartwoloops}
\left\{ l_1, -l_1 - l_2 - p_2 - p_3, l_2, l_1 + p_1 + p_2 + p_3, l_1 + p_3, l_2 + p_2, l_2 - p_1, l_2 + p_2 + p_3, l_1 + p_2 + p_3 \right\}
\end{equation}
That is, any inner product within the set of momenta $\{l_1, l_2, p_1,p_2,  p_3 \}$ is a linear combination of the listed propagators and $s$ and $t$, and can thus be used in standard IBP reduction programs. For this particular problem, LiteRed was used running on a laptop due to ease of implementation, but the problem is so simple by modern IBP standards that probably any implementation of IBP solving will yield an answer in finite time.
 
Beyond IBP reduction it is also practically advantageous to express all inner products in terms of the above set of propagators. In essence this set of propagators for all inner products solves momentum conservation on the graph level. This make implementing the cut conditions straightforward as in these coordinates this is a linear constraint. 
\begin{figure}\centering
\includegraphics[scale=0.5]{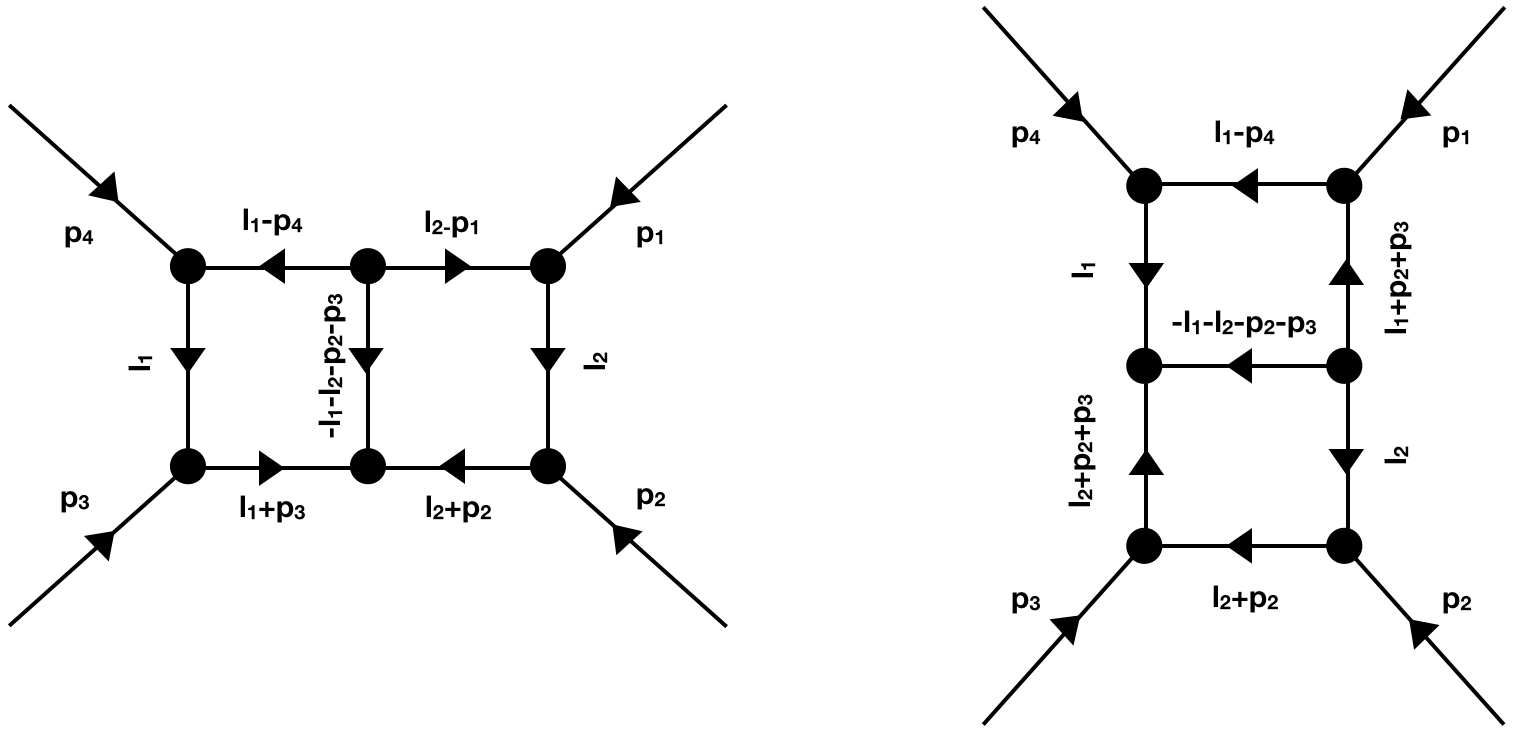}
\caption{Two double box integrals parametrised by the set in equation \eqref{eq:integralbasisplanartwoloops}.}
\label{fig:doubleboxes}
\end{figure}
The set thus obtained is special as it turns out to parametrise both planar double boxes which appear for the colour-ordered planar $\mathcal{N}=4$ super-Yang-Mills amplitude, see figure \ref{fig:doubleboxes}. Less obvious but still true is that they can be used to parametrise all two loop, trivalent planar graphs\footnote{This leads to the conjecture that a similar computation as above will yield a set of propagators that similarly completely parametrises all propagators of Feynman integrals in planar Yang-Mills amplitudes. At one loop this is straightforward to verify for n-point amplitudes  by starting with an n-gon scalar integral and pinching loop propagators.}. The solution of all IBP relations involving the above set with LiteRed yields a list of master integrals for the problem under study. Note that some of the masters integrals generated by LiteRed are disconnected integral topologies. An example are integrals over propagators
\begin{equation}
\sim \left[ (l_2-p_1)^2  (l_2+p_2)^2 (l_1 + p_3)^2 (l_1-p_4)^2 \right]^{-1}
\end{equation}
which is the product of two s-channel bubbles. However, there is also the product of two one-loop massless box functions. The latter must be disregarded for physics as it does not correspond to any amputated  connected graph. In addition there are three master integrals which do not have double cuts that match any unitarity-derived discontinuity. These three are scalar integrals which cannot be associated with a scalar graph. An example is the integrand
\begin{equation}
\sim \left[(l_2 + p_2 + p_3)^2 (l_1 + p_3)^2 (l_1 + p_1 + p_2 + p_3)^2 (l_2)^2 (l_1 + l_2 + p_2 + p_3)^2\right]^{-1}
\end{equation}
These integrals will also be disregarded. 

Cyclic symmetry shifting $(1 \rightarrow 2,2 \rightarrow 3, 3 \rightarrow 4, 4 \rightarrow 1)$ acts on the set \eqref{eq:integralbasisplanartwoloops}. After the cyclic shift the propagators can be put back into a form that matches the set through a shift of momentum,
\begin{equation}
l_1 \rightarrow l_1 - p_4 \qquad  l_2  \rightarrow l_2 + p_2
\end{equation}
which induces the following permutation
\begin{equation}
\left(\begin{array}{ccccccccc}
1 & 2& 3& 4& 5& 6& 7& 8& 9\\
5 & 2& 7& 1& 9& 3& 8& 6& 4
\end{array}\right)
\end{equation}
on the propagators (notation indicates leg $1$ is mapped to leg $5$, etc.). Note this is not a symmetry of the integrals: cyclic symmetry is a constraint on the physical planar amplitude. It can be checked by considering the graph polynomials that the list of propagators in \eqref{eq:integralbasisplanartwoloops} carries an order $8$ permutation group, generated by a choice of three independent permutations, say:
\begin{equation}
\left(\begin{array}{ccccccccc}
1 & 2& 3& 4& 5& 6& 7& 8& 9\\
3 & 2& 1& 6& 7& 4& 5& 9& 8
\end{array}\right)
\end{equation}
\begin{equation}
\left(\begin{array}{ccccccccc}
1 & 2& 3& 4& 5& 6& 7& 8& 9\\
1 & 2& 3& 5& 4& 7& 6& 8& 9
\end{array}\right)
\end{equation}
\begin{equation}\label{eq:symmap}
\left(\begin{array}{ccccccccc}
1 & 2& 3& 4& 5& 6& 7& 8& 9\\
8 & 2& 9& 6& 7& 4& 5& 1& 3
\end{array}\right)
\end{equation}

The calculational chain in this section will feature cut master integrals. There are two natural notions of cutting which are important to distinguish. These will be referred to as `lexicographic cuts' and `discontinuity cuts'. The first notion is implemented on the level of lists: it simply looks if a specific integral has non-zero coefficients at fixed positions in the ordered list \eqref{eq:integralbasisplanartwoloops}. For instance, one can look at all integrals which have unit indices for the first three positions. The lexicographic cut takes out these propagators identified by lexicographic position only and replaces them by delta functions. 

The discontinuity cut corresponds to the physical notion of a cut and instruct one to sum over all lexicographic cuts such that first the sum over cut propagators is a fixed sum of external momenta, and second that the total number of external legs on each cut piece is fixed. For instance, taking the sum of cut propagator momenta to be $p_2 + p_3$ and requiring two structures with five external legs to emerge fixes two possible cuts of the list  \eqref{eq:integralbasisplanartwoloops}: either take the first three propagators or take propagators $2$, $8$ and $9$. Hence, for for instance for the double box integral with propagator exponents
\begin{equation}
\{1,1,1,1,1,1,1,0,0\}
\end{equation}
(the left double box in  figure \ref{fig:doubleboxes}) there is only one discontinuity cut in  the $2,3$ channel, and the two notions of cut agree. A second example is the other double box integral  (the right double box in  figure \ref{fig:doubleboxes}) which in terms of the list above has propagator exponents
\begin{equation}
\{1,1,1,1,0,1,0,1,1\}
\end{equation}
Here, the discontinuity cut is the sum over the two possible lexicographic cuts: one dropping the first three propagators, and one dropping $2,8,9$. These two cuts are connected since the symmetry map in equation \eqref{eq:symmap} connects these cuts to each other. 

The discontinuity cut will be most important for us. The reason is physics: even for the much more simple example of $\phi^3$ theory, the unitarity cut of the two loop, four point amplitude would feature both Feynman graphs obtained by taking a discontinuity cut of the $\{1,1,1,1,0,1,0,1,1\}$ integral: both encode physical singularities. Two crucial assumptions will be made below. The first is that the discontinuity cut commutes with all IBP relations: if a relation is true for integrals, it should also hold for discontinuities of these integrals. This is known to be false for the lexicographic cut which can however be circumvented by using a subset of the IBP relations, typically giving a larger master integral basis in intermediate stages. Since the notion of discontinuity cut employed here is more physical this is not expected to be a problem here. The second is that there are no relations between cut integrals beyond those that descend from the parent integral IBP relations, as long as the integrals themselves do not vanish.

\subsection{Assembling the amplitude}

The list of physically interesting master integrals and the list of tensor structures immediately yields an Ansatz for the complete amplitude, with coefficients which are rational functions only of the external Mandelstams,
\begin{equation}\label{eq:ansatz2Loop4gluons}
A^2(1,2,3,4) = \sum_{j \in \textrm{masters}} \sum_{i \in \textrm{pol. struct.}} \alpha_{j,i}(s,t, D) \textrm{MI}_{j} B_i
\end{equation}
These coefficients $\alpha_{z,i}(s,t,D) $ must then be fixed by unitarity cuts. For these employ triple cuts, iterated double cuts as well as cyclicity. Several topologies such as the double boxes appear on several cuts. This yields an immediate cross-check of results and a valuable tool to check normalisations. 

\begin{figure}\centering
\includegraphics[scale=0.30]{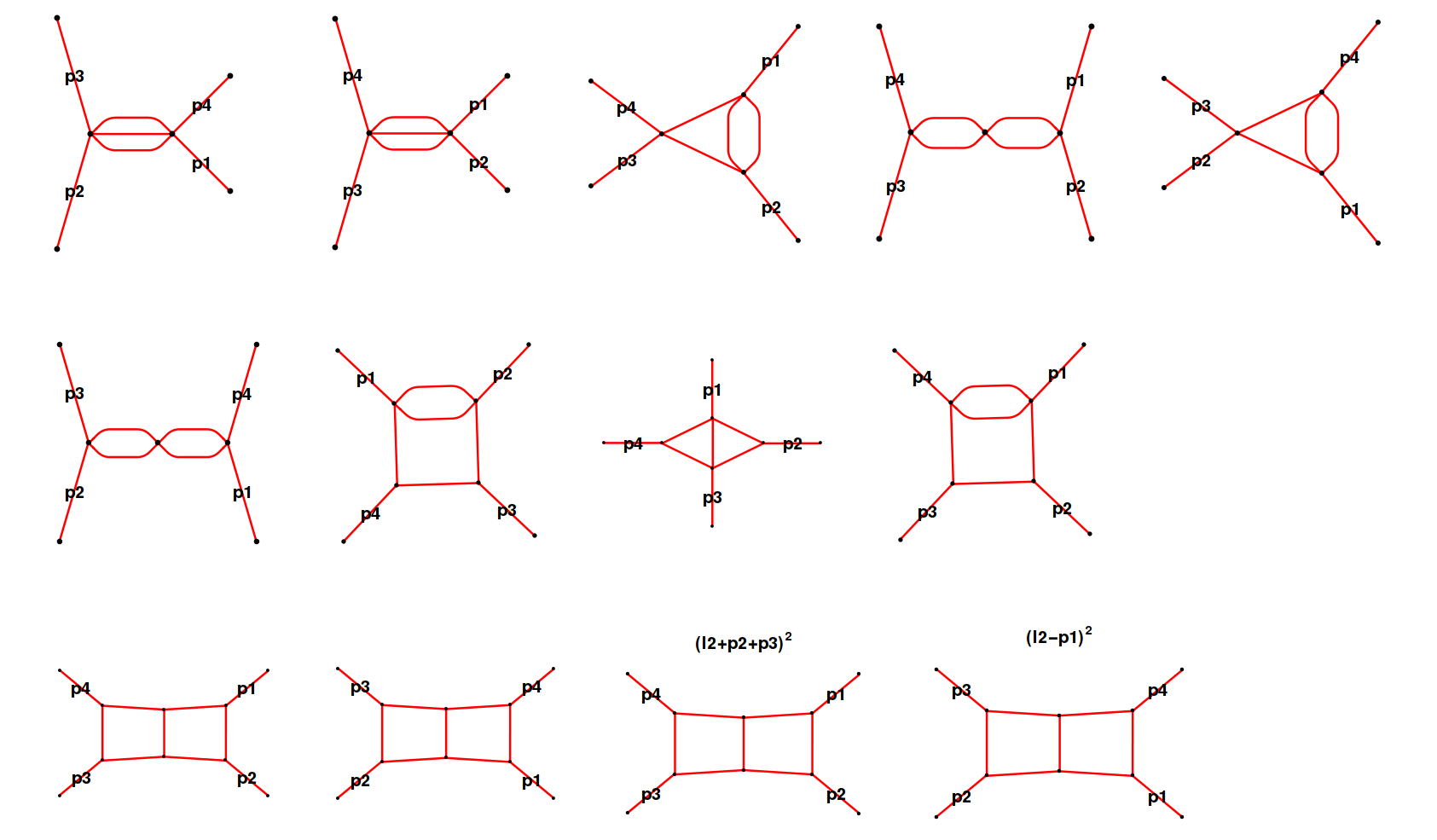}
\caption{Master integrals for four-point two-loop planar amplitude in pure Yang-Mills theory. The last two master integrals contain numerators.}
\label{fig:MIs4pt2L}
\end{figure}

What is clear from the list of master integrals (see figure \ref{fig:MIs4pt2L}) is that there are topologies which cannot fixed by the triple cut: the double bubbles in the $s$ and $t$ channels as well as the bubble replacing a propagator in a one-loop triangle. These topologies can be fixed from iterated double cuts as done here. Alternatively, one may take the one-loop four gluon amplitude and compute a single double cut of the two-loop amplitude in either $s$ or $t$ channel. 

\begin{figure}\centering
\includegraphics[scale=0.43]{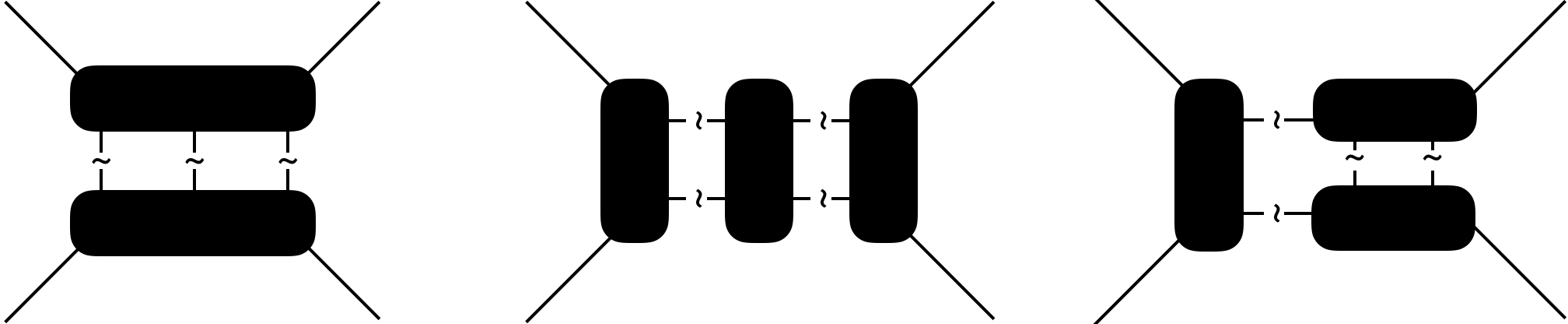}
\caption{All needed cut topologies, up to cyclicity}
\label{fig:2L4ptCuts}
\end{figure}

The calculational strategy is an extension of the one employed above in the one loop case: compute the unitarity contribution to a given cut. For this multiply the right tree amplitudes and sum over the internal helicities. Then multiply one by one with the external tensor structures and sum over external helicities. The result is an expression in terms of inner products of loop and external momenta only. Now identify the integrals and coefficients which yields the obtained expression when the discontinuity cut is applied to them. These integrals must then be IBP reduced before taking the unitarity cut. By the assumption that the cut master integrals are independent, one can then read off the coefficients of the master integrals as was done in the one loop case. 

The main difference to the one loop case is in the identification of the cut master integrals. Since the discontinuity cut can result in multiple terms per the example above, care has to be taken not to over-count. However, since the unitarity derived expression contains all tree level singularities this expression automatically contains the multiple terms: it can be checked explicitly that the symmetry map mapping one lexicographic cut to the other is a symmetry of the unitarity cut expression for the terms containing either cut. To get the correct answer, one must therefore only divide by the number of lexicographic cuts of a given type that can be taken of a particular integral after reinstating the naively cut propagators. 

As a further example of the pitfalls of discontinuity cut and lexicographic cut, consider the `crossed' iterated double cut: first in the s, then in the t channel. This leads to the cut topology on the right of figure \ref{fig:2L4ptCuts}, plus its mirror image. The sum of these two contributions is the discontinuity cut, while one of these is a lexicographic cut. Although one can use standard IBP relations for the lexicographic cut in this case, one cannot use relations derived from momenta mappings that keep the Mandelstams invariant when employing the lexicographic cut (in LiteRed, including these is a secondary option for instance). For four particle amplitudes, these maps are exactly the reflections. 

Pushing through the computation yields an answer in finite time, using only modest laptop computing resources. The result is attached to the arXiv submission of this article, where the integrals involve reverse lexicographic order compared to the list \eqref{eq:integralbasisplanartwoloops} for idiosyncratic reasons. New compared to the one loop computation is the need for multiple cuts to fix all coefficients. The triple cut yields most integral coefficients, except for the triangle-bubble and the `glasses' type topologies. These have to be fixed by iterated double cuts. The unitarity cuts must be consistent with cyclic symmetry. For the used four gluon tensor structure basis the first seven elements are cyclic themselves, so consistency can be checked directly on the coefficients. For the three non-symmetric elements the non-trivial transformations of the tensor structures has to be taken into account in principle. Since these coefficients vanish already before reconstruction to the amplitude basis this is not an issue here.

\subsubsection*{Need for speed}
The feasibility of the calculational approach depends on the speed of the implementation. At one loop this is not that much of an issue for gluons, but at two loops some care is necessary. This subsubsection will list some tips and tricks. 

A recurring problem is that the sum over helicities generates spurious dependence on a gauge vector in intermediate expressions. For small to intermediate expressions showing cancellation explicitly is possible, but for large expressions such as the ones generated in the two-loop computation of this section this becomes hard without invoking special computing resources. To battle this one should use the same gauge vector $q$ for all polarisation vectors in the problem. Furthermore, one should define in Mathematica
\begin{equation}
\xi_i \cdot q = 0
\end{equation}
for all polarisation vectors to prevent intermediate expression swell. To obtain further simplification of an expression known to be algebraically independent of a vector one can, \emph{after} summing over all helicities, set $q=p_1-p_2$ as this reduces the number of independent letters in a large expression. To check gauge invariance compute the difference of setting $q=p_1-p_2$ to, say, setting $q=p_3-p_2$. Moreover, often the only source of poles in intermediate expressions is the denominator in the polarisation state sum and therefore checking absence of poles equates to checking gauge invariance in this case. Checking gauge invariance in intermediate stages is highly recommended as a consistency check.

A major speed bottleneck is the projection of the cut loop amplitude expression onto the basis of external kinematic factors. Speed can be obtained by splitting the computation into two parts. First, construct all possible contractions of external polarisations and internal and external momenta subject to transversality and momentum conservation, but \emph{not} on-shell gauge invariance. A given cut loop amplitude expression can be expressed in this set using linear algebra approach, which is very fast. On the other hand one can project this set onto the external kinematic basis. This second step is by itself gauge-variant, but relatively fast. Combining computations yields the desired, gauge-invariant answer. A major additional advantage is that the second step only has to be done once. This intermediate result can then be used for all cuts of two-loop, four gluon amplitudes with arbitrary internal matter content, for general effective field theories.

\subsection{Results in 't Hooft-Veltman scheme}

To compare to the known results in \cite{Bern:2000dn} we have performed the computation in 't Hooft-Veltman (HV) scheme. Much of the computation is the similar to the previous case, such as the propagators of eq.(\ref{eq:integralbasisplanartwoloops}), and  the structure of the Ansatz made in eq.(\ref{eq:ansatz2Loop4gluons}).  The difference of HV scheme to the all-D-dimensional scheme appears while projecting the results after summing over the internal helicities to the external tensor structures.

Since the internal particles are living in D dimensions while the external ones are in four dimensions, the external helicity summation yields $\mu$-type terms. Decompose the D-dimensional internal momenta into $4$ plus $D-4$ dimensional components:
\bea
&&\,l_1 = \hat l_{1}+\vec \mu_1,\qquad \qquad\quad    l_2 = \hat l_{2} +\vec \mu_2,\\
&&(l_1)^2 = (\hat l_{1})^2-\mu_1^2,\qquad   (l_2)^2 = (\hat l_{2})^2 -\mu_2^2
\eea
where $\hat l_{1}$ and $\hat l_{2}$ are four dimensional components, and $\vec \mu_1$ and $\vec \mu_2$ are $D-4$ dimensional components. 

With Minkowski metric $(+1, -1, -1, \dots)$, the inner products of D-dimensional loop momenta are decomposed as in the second line, where $\mu_1^2= \vec \mu_1\cdot \vec \mu_1$, $\mu_2^2= \vec \mu_2\cdot \vec \mu_2$ and $\mu_{12}^2=(\vec \mu_1+\vec \mu_2)^2$. The inner products after summing over external helicities are only of four dimensional components, i.e., $\hat l_1^2$, $\hat l_1\cdot \hat l_2$, $\hat l_2^2$ and $\hat l_{1(2)}\cdot p_i$. They can be expressed in terms of the propagators in eq.(\ref{eq:integralbasisplanartwoloops}) together with $\mu_1^2$, $\mu_2^2$ and $\mu_{12}^2$.

As above, identify the integrals and coefficients which, when the discontinuity cut is applied to them yields the obtained expression. The integrals now contain certain powers of $\mu_1^2$, $\mu_2^2$ and $\mu_{12}^2$ and their products. These $\mu^2$ terms are first translated into the higher dimensional integrals, using the approach introduced in section 4.1 in \cite{Bern:2002tk}. LiteRed \cite{Lee:2013mka, Lee:2012cn} can compute the dimensional recurrence relations, which lower the integrals from higher dimension $D+2 n$ to $D$ \citep{Tarasov:1996br}. Once all integrals are lowered into $D$ dimension, they can be IBP reduced to master integrals. The coefficients of the master integrals containing respective unitarity cuts can be read off. The last step is the same as all-D-dimensional scheme. 

\begin{figure}\centering
\includegraphics[scale=0.30]{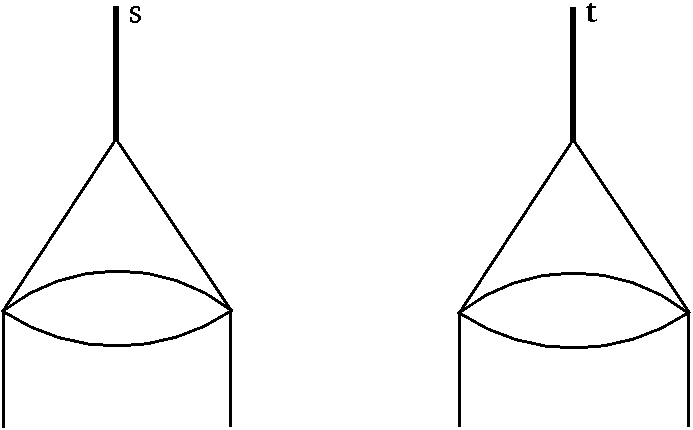}
\caption{The master integrals determined by crossed double cut: the last cut way in figure \ref{fig:2L4ptCuts} }
\label{fig:master-integrals-cross-cuts}
\end{figure}
To explore coefficients of master integrals where a bubble replacing a propagator in a one-loop triangle  shown in figure \ref{fig:master-integrals-cross-cuts}, iterative cross cuts (last one in figure \ref{fig:2L4ptCuts}) can be applied. Consider the left master integral first, two ways can contribute its coefficients: (1) cut two legs attaching to $3$ and $4$ first, and then the middle leg and the leg between $1$ and $2$; (2) cut two legs attaching to $1$ and $2$ first, then the middle and the one between $3$ and $4$. The integrals which come from these two types of cuts and contain both cut-legs should be multiplied a factor $1/2$ to avoid over counting. This also applies to the right master integral.

In the HV scheme the coefficients of the non-symmetric tensor structures vanish just as they did in the D-dimensional scheme. The identical helicity results match those in \cite{Bern:2000dn} precisely. An advantage of this construction strategy is that it is straightforward to derive all different helicity configuration amplitudes with one computational step. 

\section{Beyond one loop: generalities}
\label{sec:morethanoneloop}
This section collects general results on computations beyond the one loop order. 

\subsection{Sewing relations for iterated double cuts}
In $\mathcal{N}=4$ SYM and $\mathcal{N}=8$ SUGRA theories there are graphics-based rules which are routinely used to fix the coefficients of a subset of the master integrals by automatically taking into account iterated double cuts, to the loop order required.  The techniques presented above allow for a generalisation of this rule to non-supersymmetric theories. Even more generally, in this section it is shown how to derive sewing relations: given two four point amplitudes at fixed loop order, one can compute the contribution of the double cut where these quantities appear on each side. This generalisation applies to all fixed loop order computations, greatly simplifying some of the needed algebra. 

The computation is a generalisation of the one loop case. As input express two amplitudes in the kinematic basis,
\begin{align}
A_{\textrm{left}}(l_1,1,2,l_2) & =  \sum_{i} \alpha_i B_i (s, t_L), \\
A_{\textrm{right}}(-l_2,3,4,-l_1) & =  \sum_{i} \beta_i  B_j (s, t_R), 
\end{align}
with $t_L =(l_1+p_1)^2$ and $t_R = (l_2-p_3)^2$.
Now compute the projection of the cut onto the same kinematic basis and extract the $n$-th coefficient,
\begin{equation}
\left(P^{-1}\right)_{n}^{\,\,\,k} \sum_{\textrm{all helicities}} B_k (s,t) A_{\textrm{left}}(l_1,1,2,l_2) A_{\textrm{right}}(-l_2,3,4,-l_1).
\end{equation}
The point of this seemingly complicated-looking expression is that it only involves tree-level quantities: this computation can be done once and for all. A particularly useful special case of this formula involves a tree amplitude on one of the sides.  Given a scattering amplitude at $L$ loops
\begin{align}
A^{L}(l_1,1,2,l_2) & =  \sum_{i} \alpha^L_i(s,t_L) B_i (s, t_L),
\end{align}
one can use this formula to compute the coefficients on the cut where a four point tree level amplitude is factored off by matrix multiplication
\begin{equation}
\textrm{Cut}\,\, \alpha^{L+1}_i(s,t)  = \left(P^{-1}\right)_{i}^{\,\,\,j} \sum_{\textrm{all helicities}} B_j (s,t) A^L(l_1,1,2,l_2) \, A^0 (-l_2,3,4,-l_1).
\end{equation}
This is the sought-for generalisation of the rung rule. In particular, it shows that for this class of iterated cuts one can compute sums over internal helicities efficiently, if the tree level problem contained in the last equation can be solved. However, the appearance of non-local poles in the four point amplitudes at one loop obtained above (typically with vanishing residue) show that the obtained representation may need more work to be effective in computations.

\subsection{Colour algebra using projectors}
In general, in a theory with adjoint matter the colour factors at higher loop orders are combinations of structure constants. Since there is no known all-multiplicity equivalent to the Del-Duca-Dixon-Maltoni \cite{DelDuca:1999rs} basis above one loop order, the colour quantum numbers require some care. Minimal solutions of the colour-Jacobi identities at higher loop order have been studied, see e.g. \cite{Naculich:2011ep,Edison:2012fn, reichenbaecherthesis13, Ochirov:2016ewn}. These typically involve explicitly solving the colour-Jacobi identities for all trivalent graphs which can be time-consuming, tedious and error-prone. Here a different approach is introduced which has the advantage of easy (computer) implementation. The approach is the direct colour-algebra analogue of the approach to the kinematic analysis pursued above. 

Given a basis $B_i$ of colour structures for a given problem one typically needs to compute the projection coefficients 
\begin{equation}\label{eq:colexpcoef}
f[a_1,a_2,\ldots, a_n] = \sum_i \alpha_i B_i[a_1,a_2,\ldots, a_n] 
\end{equation}
for a given basis set of colour structures $f$. For this one multiplies left and right by a basis element $B_j$,
\begin{equation}
B_j[a_1,a_2,\ldots, a_n] f[a_1,a_2,\ldots, a_n] = \sum_i \alpha_i B_i[a_1,a_2,\ldots, a_n] B_j[a_1,a_2,\ldots, a_n]
\end{equation}
and sums over all colour indices left and right. This leads to a linear equation
\begin{equation}
f_j  = \sum_i \alpha_i P_{ij}
\end{equation}
where 
\begin{equation}
f_i = \sum_{colour} f[a_1,a_2,\ldots, a_n] B_i[a_1,a_2,\ldots, a_n]
\end{equation}
and
\begin{equation}\label{eq:colourprojectionmat}
P_{ij} = \sum_{colour}B_i[a_1,a_2,\ldots, a_n]B_j[a_1,a_2,\ldots, a_n]
\end{equation}
Since $B$ is a basis, the matrix $P$ is invertible and the matrix coefficients are obtained. Computation of the expansion coefficients is thus reduced to computing complete contractions of all colour indices. These amount to vacuum diagrams in colour space.  

The problem of computing exactly these vacuum diagrams in colour space has essentially been solved for at least all cases of collider relevant interest in \cite{vanRitbergen:1998pn} and implemented in a publicly available FORM package. For those of us not that conversant with FORM a simple interface script to Mathematica can be constructed for convenience. This reduces the computation to simple and fast manipulations. Although this will not be pursued in this article, one can use this type of technology also to systematically obtain explicit basis sets by generating all trivalent graphs and dressing these with colour structures: relations between colour structures show up in the kernel of the would-be projection matrix computed as in \eqref{eq:colourprojectionmat}. In general one can obtain relations with coefficients which depend on general group theory invariants. See also \cite{Boels:2012ew} for an overview over invariants which can appear in particular contractions of the structure constants in the adjoint and their values for the algebra $SU(N)$. 

The technique introduced here generalises easily to more general representations than just the adjoint. This includes especially the fundamental representation which is of immediate physical interest. What changes is the set of basis elements, which now has to include more general indices and has to be determined on a case-by-case basis. The computation still reduces to the computation of vacuum graphs, for which the needed colour algebra is also covered by the techniques and, for the fundamental, by the FORM implementation of \cite{vanRitbergen:1998pn}. Note that the running time will increase if fundamental representation matrices are included. For all practical physical purposes, the approach advocated here solves the colour-algebra once and for all. A choice of colour basis is needed as an important milestone on the road to a practical implementation. As a final remark, the coefficients in the expansion \eqref{eq:colexpcoef} in general are dependent on group data such as $N_c$ for $SU(N_c)$, encoding for instance sub-leading colour information.

\section{Discussion and conclusions}

In this article we have introduced a general framework to compute a class of quantum field theory observables, using nothing but Poincar\'e symmetry, locality and unitarity. Several example computations with mixed field content show the applicability of the method, which is especially interesting at the quantum level. As stated in the introduction, several elements of the method have been introduced separately before. This includes solving IBP relations, the power of unitarity to constrain complete scattering amplitudes as well as the importance of keeping track of tensor structures. What is especially new here is the combination of the use of tensor structure variables for scattering amplitudes with unitarity and IBP relations on the cuts. This allows one to reduce the cut amplitudes to cut scalar integrals. In turn this can then be used to deploy fairly standard technology of IBP relations. For the later the relation between cut integral and cutting amplitudes is important as is shown in the two loop example of the previous section. For speed it would be better to solve IBP relations using the information on the cut, for instance as is done in \cite{Abreu:2017xsl}. To our knowledge all current IBP relation solving codes employ the lexicographic cut condition. It would certainly be interesting to explore this condition as it is easier to implement; here we opted for mostly using the more physical notion of discontinuity cut. From a foundational point of view it would be interesting to prove that loop amplitudes are determined completely by tree amplitudes. As an operative assumption it is physically reasonable, and certainly bears out in the computations presented here. This foundational question intersects non-trivially with the study of IR and UV divergences.  

The substructure uncovered in the amplitudes calculated above such as the vanishing of certain kinematic basis coefficients deserves further study. Crucial would be to study the extension to higher points: this would allow one to decide if the vanishing is an artefact of four point kinematics or the result of a more basic symmetry of nature. More generally the techniques introduced here can at least in principle be applied widely. Immediate targets include planar and non-planar gluon amplitudes as well as amplitudes which contain gravitons with more legs and/or loops. The loop direction seems easier as one only has to compute for instance the inverse of the relevant P-matrix once. Amplitudes in effective field theories can be computed in principle straightforwardly. The use of tree amplitudes to constructs the cuts of loop amplitudes significantly simplifies computations compared to Feynman graph methods. This also allows one to tap into the extensive literature on the structure of tree amplitudes. The integrand is naturally reduced in complexity by the use of BCJ relations for instance. The use of other representations of tree amplitudes on the cuts such as those in \cite{Cachazo:2013hca} should be explored: the Pfaffians of that method are after all a particular choice of solution to the on-shell constraints.  

The inclusion of massless and massless fermions is certainly within immediate reach \cite{upcomingMass}. For the massless case, see also \cite{Glover:2003cm}. We would like to point out to also more general observables such as form factors can immediately and naturally be included in the above framework. Gauge invariant operators are modelled in this framework as massive, mostly scalar legs, potentially with internal charge assignments. The application of P-matrices to analyse colour offers a practical solution to vexing practical problems especially for computing non-planar quantities: finding a colour-basis, relating a given structure to the basis and keeping track of signs. This certainly could and should be extended much further. A further target would be to study theories with massless higher spin matter, see for instance \cite{Roiban:2017iqg} for work in this direction. Similarly, since the presented techniques are so general it would be interesting to see what can be learned about the structure of quantum field theory. 

An interesting feature of the results of the computations performed here is that generically the scattering amplitudes at loop level will include high order apparent kinematic divergences, such as the $\frac{1}{u^4}$ type pole in the planar Yang-Mills amplitudes as the coefficient of box and bubble amplitudes. The cancellation of the residues of this pole down to finite, $~u^0$ terms is a requirement of locality which we have verified for one loop amplitudes. Note this requirement is exact in $D$ dimension. The cancellations involve the IBP master integrals explicitly, which suggests these poles are a consequence of a bad choice of master integral basis. It would be very interesting if a more physical basis of master integrals exists in general. Furthermore, systematising the constraints from tree level unitarity on general integral topologies should be fruitful. In the presented method the origin of the unphysical poles of gluon amplitudes is manifest: they arise from inverting the P-matrix. Hence it is expected that the apparent unphysical poles are a generic feature of any computation of the non-supersymmetric scattering amplitudes. An interesting further perspective is opened by the consideration of IR divergences in the approach introduced here. 

The methods of this article apply most naturally to non-supersymmetric scattering amplitudes in dimensional regularisation.  As such the approach naturally extends to the study of full cross-sections. This unlocks a door toward physical applications such as to computations within the Standard Model of particle physics.

\acknowledgments
It is a pleasure to thank Nima Arkani-Hamed, Sven-Olaf Moch, Qingjun Jin, Frederik Bartelmann and Yang Zhang for discussions. This work was supported by the German Science Foundation (DFG) within the Collaborative Research Center 676 ``Particles, Strings and the Early Universe''.

\appendix

\section{Spinor helicity expressions for basis elements and their relations}
In this section, explicit expressions for four-gluon basis included in auxiliary files in terms of standard four dimensional spinor helicity formalism and Mandelstam variables are given.
Denote these ten basis elements as $B_i$ with $(i=1,\cdots,10)$, in which ${B_1, \dots, B_7}$ are those of Bose symmetry while ${B_8, \dots, B_{10}}$ are anti-symmetric of particles $(2,3,4)$.
The mass dimension of these ten basis elements are $dim=\{4,4,6,6,6,8,8,8,10,12\}$ respectively.
%Write symmetric structures of Mandelstam variables as 
%\bea
%\sigma_1={1\over 4} (s^2+t^2+u^2), \qquad \sigma_2={1\over 8} s\,t\,u
%\eea
%and the anti-symmetric variables structures of Mandelstam variables as
%\bea
%\sigma_3={1\over 8} (s-t)(t-u)(u-s)
%\eea

%%%%%%%%%
\subsection{Spinor helicity expressions}

\subsubsection*{All plus}
The explicit expressions for the basis of all plus type are
\bea
&&B_1=\,\frac{16\, s\,t \,(s^2+t^2+u^2)}{\langle 1,2\rangle \langle 2,3\rangle \langle 3,4\rangle \langle 4,1\rangle}, 
\qquad B_3=-\,\frac{12\, s^2\,t^2 u}{ \langle 1,2\rangle \langle 2,3\rangle \langle 3,4\rangle \langle 4,1\rangle}, \\
&&B_4=\,\frac{16\,s^2\,t^2 u}{ \langle 1,2\rangle \langle 2,3\rangle \langle 3,4\rangle \langle 4,1\rangle}, 
\qquad B_5=\,\frac{4\,s^2\,t^2 \,u}{ \langle 1,2\rangle \langle 2,3\rangle \langle 3,4\rangle \langle 4,1\rangle},\\
&&B_7=\, \frac{ 2\,s\,t \,(s^2+t^2+u^2)^2}{\langle 1,2\rangle \langle 2,3\rangle \langle 3,4\rangle \langle 4,1\rangle}, 
\qquad B_2=B_6=B_8=B_9=B_{10}=0.
\eea

%%%%%%%%%%%%%%%%%
\subsubsection*{Single minus}
The single minus case contains four distinct helicity configurations: one for each minus at a particular particle. The former seven Bose-symmetric basis elements for these four single minus helicity configurations are the same up to certain prefactors $H_{\rm{SM}}$ which contain information of helicity configurations.
The biggest difference emerges in the last three basis elements $B_i,~i=8,9,10$, which come from anti-symmetric permutations of $(2,3,4)$.
\bea\label{SpinHelicity-SM}
&&B^{\rm{SM}}_4=\, B^{\rm{SM}}_5=4\,s^2\,t^2\,H_{\rm{SM}},\\
&&B^{\rm{SM}}_1=B^{\rm{SM}}_2=B^{\rm{SM}}_3=B^{\rm{SM}}_6=B^{\rm{SM}}_7=0.\\
\eea
where the prefactors $H_{\rm{SM}}$ for different helicity configurations are 
\bea
H_{\rm{SM}}(1^-,2^+,3^+,4^+)=\frac{[2,4]^2}{ [1,2]\langle 2,3\rangle \langle 3,4\rangle [4,1]},\\
H_{\rm{SM}}(1^+,2^-,3^+,4^+)=\frac{[1,3]^2}{ [2,3]\langle 3,4\rangle \langle 4,1\rangle [1,2]},\\
H_{\rm{SM}}(1^+,2^+,3^-,4^+)=\frac{[2,4]^2}{ [3,4]\langle 4,1\rangle \langle 1,2\rangle [2,3]},\\
H_{\rm{SM}}(1^+,2^+,3^+,4^-)=\frac{[1,3]^2}{ [4,1]\langle 1,2\rangle \langle 2,3\rangle [3,4]}.
\eea

The last three basis elements of single minus type are expressed respectively as 
\begin{itemize}

\item $(1^-,2^+,3^+,4^+)$
\bea
&&B^{\rm{SM}}_8(1^-,2^+,3^+,4^+)=0,\\
&&B^{\rm{SM}}_9(1^-,2^+,3^+,4^+)=0,\\
&&B^{\rm{SM}}_{10}(1^-,2^+,3^+,4^+)=-2s^2\,t^2\,(s-t)\,(t-u)\,(u-s)\,H_{\rm{SM}}(1^-,2^+,3^+,4^+).
\eea

\item $(1^+,2^-,3^+,4^+)$
\bea
&&B^{\rm{SM}}_8(1^+,2^-,3^+,4^+)=-8\,s^2\,t^2\,(t-u)\,H_{\rm{SM}}(1^+,2^-,3^+,4^+),\\
&&B^{\rm{SM}}_9(1^+,2^-,3^+,4^+)=4\,s^3\,t^2\,(t-u)\,H_{\rm{SM}}(1^+,2^-,3^+,4^+),\\
&&B^{\rm{SM}}_{10}(1^+,2^-,3^+,4^+)=-s^2\,t^2\,(t-u)\,(3\,s^2+t^2+u^2)\,H_{\rm{SM}}(1^+,2^-,3^+,4^+).
\eea

\item$(1^+,2^+,3^-,4^+)$ 
\bea
&&B^{\rm{SM}}_8(1^+,2^+,3^-,4^+)=-8\,s^2\,t^2\,(s-t)\,H_{\rm{SM}}(1^+,2^+,3^-,4^+),\\
&&B^{\rm{SM}}_9(1^+,2^+,3^-,4^+)=4\,s^2\,t^2\, u\,(s-t)\, H_{\rm{SM}}(1^+,2^+,3^-,4^+),\\
&&B^{\rm{SM}}_{10}(1^+,2^+,3^-,4^+)=-s^2\, t^2\,(s-t)\, (s^2+t^2+3\,u^2)\,H_{\rm{SM}}(1^+,2^+,3^-,4^+).
\eea

\item $(1^+,2^+,3^+,4^-)$
\bea
&&B^{\rm{SM}}_8(1^+,2^+,3^+,4^-)=-8\,s^2\,t^2\,(u-s)\,H_{\rm{SM}}(1^+,2^+,3^+,4^-),\\
&&B^{\rm{SM}}_9(1^+,2^+,3^+,4^-)=4\,s^2\,t^3\, (u-s)\,H_{\rm{SM}}(1^+,2^+,3^+,4^-),\\
&&B^{\rm{SM}}_{10}(1^+,2^+,3^+,4^-)=-s^2\,t^2\,(u-s)\,(s^2+3\,t^2+u^2)\,H_{\rm{SM}}(1^+,2^+,3^+,4^-).
\eea

\end{itemize}
%%%%%%%%

\subsubsection*{MHV}

Helicity configuration $(1^-,2^-,3^+,4^+)$ and its conjugate $(1^+,2^+,3^-,4^-)$ share the same kinematic kernel. After multiplying with corresponding MHV tree amplitudes which contain the helicity information, complete spinor helicity expressions are obtained. This also happens for $(1^+,2^-,3^-,4^+)$ and its conjugate $(1^-,2^+,3^+,4^-)$ as well as non-adjacent MHV configuration $(1^-,2^+,3^-,4^+)$ and its conjugate $(1^+,2^-,3^+,4^-)$.

In every MHV case, $B_4= B_5= B_6= B_8= B_9= B_{10}=0$.

\begin{itemize}
\item For helicity $(1^-,2^-,3^+,4^+)$ and its conjugate $(1^+,2^+,3^-,4^-)$, 
\bea
&&B_1=64 \, s\,t\,A(1^\mp,2^\mp,3^\pm,4^\pm) , \qquad ~~ B_2=16\, s\,t \, A(1^\mp,2^\mp,3^\pm,4^\pm),\\
&&B_3=-4\,s^2\,t \, A(1^\mp,2^\mp,3^\pm,4^\pm),
\qquad B_7=4\,s^3\,t\,A(1^\mp,2^\mp,3^\pm,4^\pm),
\eea
where $A(1^\mp,2^\mp,3^\pm,4^\pm)$ denotes tree amplitudes of $(1^-,2^-,3^+,4^+)$ and $(1^+,2^+,3^-,4^-)$.

\item For helicity $(1^+,2^-,3^-,4^+)$ and its conjugate $(1^-,2^+,3^+,4^-)$, 
\bea
&&B_1=64 \, s\,t\,A(1^\pm,2^\mp,3^\mp,4^\pm) , \qquad ~~ B_2=16\, s\,t \, A(1^\pm,2^\mp,3^\mp,4^\pm),\\
&&B_3=-4\,s\,t^2 \, A(1^\pm,2^\mp,3^\mp,4^\pm),
\qquad B_7=4\,s\,t^3\,A(1^\pm,2^\mp,3^\mp,4^\pm),
\eea
where $A(1^\pm,2^\mp,3^\mp,4^\pm)$ denotes tree amplitudes of $(1^+,2^-,3^-,4^+)$ and $(1^-,2^+,3^+,4^-)$.

\item For helicity $(1^-,2^+,3^-,4^+)$ and its conjugate $(1^+,2^-,3^+,4^-)$,
\bea
&&B_1=64 \, s\,t\,A(1^\mp,2^\pm,3^\mp,4^\pm) , \qquad ~~ B_2=16\, s\,t \, A(1^\mp,2^\pm,3^\mp,4^\pm),\\
&&B_3=-4\,s\,t\,u \, A(1^\mp,2^\pm,3^\mp,4^\pm),
\qquad B_7=4\,s\,t\,u^2\,A(1^\mp,2^\pm,3^\mp,4^\pm),
\eea
where $A(1^\mp,2^\pm,3^\mp,4^\pm)$ denotes tree amplitudes of $(1^-,2^+,3^-,4^+)$ and $(1^+,2^-,3^+,4^-)$.
\end{itemize}

%%%%%%%
\subsection{Relations among the basis elements in four spacetime dimensions}\label{sec:relations}

The basis element $B_6$ vanishes for all helicity configurations, which means it does not contribute to any amplitude constructions in four dimensional spacetime. In other words, it is evanescent. 

To check if further relations occur among the basis elements in four dimensions, the basis elements need to be rearranged into uniform mass dimensional first by multiplying proper polynomials of Mandelstam variables. 
In this way, the basis set is enlarged. For instance, to obtain extended basis elements with uniform mass dimension $12$, original selected basis elements, of mass dimensions $\{4,4,6,6,6,8,8,10,12\}$ respectively, are multiplied by all possible $s^i\,t^j$ with $2(i+j)= 12-\textrm{dim}(B_i)$.   
Then, by taking different random integers for the enlarged basis set and joining them into a matrix, the null space of this matrix can tell the relations among the enlarged basis set.
The streamlined relations are finally derived by dropping redundant ones.

The above way can be applied to helicity configurations one by one, which gives relations for every single helicity configuration. 
It can be also applied to all helicities at the same time, then the complete relations for the full basis elements are obtained.

The following relations are found following this way, in particular, the complete relations for all helicities agree with those by studying the null space of the P-matrix in four-dimensional spacetime.

\subsubsection*{Relations for all plus and single minus helicities}
First consider basis elements in all plus and single minus helicity configurations, which vanish in theories with unbroken supersymmetry and a conserved $U(1)_R$ symmetry.
\begin{itemize}
\item With all plus helicities, four relations are derived
\bea
B_3+\frac{3\,s\,t\,u}{ 4\,(s^2+t^2+u^2)}B_1 =0\,,\,\qquad B_4-\frac{\,s\,t\,u}{ (s^2+t^2+u^2)}B_1 =0\,,\\
B_5-B_4-B_3 =0\,, \qquad B_7-\frac{1}{ 8} (s^2+t^2+u^2) B_1 =0\, .
\eea
\item With single minus helicities, one relation is found $B_5-B_4=0$. 

Note that according to eqs.(\ref{SpinHelicity-SM}), $B_1$, $B_2$, $B_3$, $B_6$, $B_7$ are all zero, namely, any relation if only including them still holds. See the succeeding relations shown below.  
\item Combining all plus and single minus helicities there are three relations
\bea
B_3+\frac{3\,s\,t\,u}{ 4\,(s^2+t^2+u^2)}B_1 =0,\qquad B_5-B_4-B_3 =0, \qquad B_7-\frac{1}{ 8} (s^2+t^2+u^2) B_1 =0\nonumber\\
\eea
\end{itemize}

\subsubsection*{Relations for MHV helicities}
Now consider the basis elements in MHV helicity configurations which can be supersymmetrized.

\begin{itemize}
\item With adjacent MHV helicities, e.g. $(-,-,+,+)$, two relations are found\\
$\,\,4B_2-B_1=0\, , \quad 16 B_7-16\,u\,B_3+s\,t\,B_1=0$.
\item With non-adjacent MHV helicities, e.g. $(-,+,-,+)$, three relations are found \\ 
$\,\,4B_2-B_1=0\, ,\quad 16B_3+uB_1=0\,,\quad  B_7+u\,B_3=0$.
\item Combine two MHV results, only one relation $\,\, 4\, B_2-B_1=0$ is obtained.
\end{itemize}

\subsubsection*{Complete relations for all helicity configurations}
If all helicities join into considerations, one relation is found: 
\bea
3 \,s \,t \,u (B_1 - 4 B_2) - 4 (s^2 + t^2 + u^2) (B_4 - B_5) = 0
\eea
It can be checked directly that this relation agrees with the one derived from the P-matrix.

\bibliographystyle{JHEP}

\bibliography{References}

\end{document}